\pdfoutput=1 
\documentclass[]{SciPost}

\hypersetup{
    colorlinks,
    linkcolor={red!50!black},
    citecolor={blue!50!black},
    urlcolor={blue!80!black}
}

\usepackage[bitstream-charter]{mathdesign}
\DeclareSymbolFont{usualmathcal}{OMS}{cmsy}{m}{n}
\DeclareSymbolFontAlphabet{\mathcal}{usualmathcal}

\fancypagestyle{SPstyle}{
\fancyhf{}
\lhead{\colorbox{scipostblue}{\bf \color{white} ~SciPost Physics }}
\rhead{{\bf \color{scipostdeepblue} ~Submission }}

\fancyfoot[C]{\textbf{\thepage}}
}

\usepackage[notref,final]{showkeys}

\usepackage{graphicx}
\usepackage{bm}

\usepackage{dcolumn}
\newcolumntype{.}{D{.}{.}{-1}}
\newcolumntype{d}[1]{D{.}{.}{#1}}

\DeclareMathOperator{\Tr}{Tr}

\DeclareMathOperator{\arccosh}{arccosh}
\DeclareMathOperator{\sgn}{sgn}
\DeclareMathOperator{\diag}{\mathbf{diag}}

\newcommand{\bS}{\begin{subequations}}
\newcommand{\eS}{\end{subequations}}

\newcommand{\CC}[1]{{}}
\newcommand{\XX}[1]{{}}

\newcommand{\fred}[1]{{\color{blue}#1}}
\newcommand{\luca}[1]{{\color{purple}#1}}
\newcommand{\oo}[1]{{\color{green}#1}}

\newcommand{\notused}[1]{{\color{gray}#1}}

\newcommand{\mat}[1]{\bm{\mathrm{#1}}}
\DeclareMathAlphabet{\mathsf}{OT1}{\sfdefault}{m}{n}
\SetMathAlphabet{\mathsf}{bold}{OT1}{\sfdefault}{bx}{n}
\newcommand{\bmat}[1]{\bm{\mathsf{#1}}}

\newcommand{\LL}{{\leftarrow\mkern-21mu\rightarrow}}
\newcommand{\MM}{{\updownarrow}}
\newcommand{\T}{{\scriptscriptstyle\mathsf{T}}}
\newcommand{\dd}{\mathrm{d}}
\newcommand{\ee}{\mathrm{e}}
\newcommand{\ii}{\operatorname{i}}

\newcommand{\Ut}{\bmat{U}^\MM}
\newcommand{\Uz}{\bmat{U}^\LL}

\newcommand{\MB}[1][]{\mathnormal{#1 M}_\mathrm{B}}
\newcommand{\mB}[1][]{\mathnormal{#1 m}_\mathrm{B}}

\newcommand{\st}{\uparrow\!\downarrow}

\newcommand{\EX}[1][]{{{
\mathrm{(ex#1)}}}}

\newcommand{\OO}   {{{
\mathrm{(o,o)}}}}
\newcommand{\OST}  {{{
\mathrm{(o,\st)}}}}
\newcommand{\OP}   {{{
\mathrm{(o,+)}}}}
\newcommand{\OK}   {{{
\mathrm{(o,\kappa)}}}}
\newcommand{\OR}   {{{
\mathrm{(o,r)}}}}
\newcommand{\OmB}  {{{
\mathrm{(o,\mB=0)}}}}
\newcommand{\OmBdv}{{{
\mathrm{(o,\mB=3/4)}}}}

\newcommand{\fb}{{{
f_\mathrm{b}}}}
\newcommand{\fs}[1][]{{{
f_\mathrm{s}^\mathrm{#1}}}}
\newcommand{\fsp}{\fs[(+)]}
\newcommand{\fsst}{\fs[(\st)]}
\newcommand{\fsr}[1][(r)]{{{
\bar f_\mathrm{s}^\mathrm{#1}}}}

\newcommand{\fsrHLN}  {0.26536(96\pm10)}
\newcommand{\fsrISN}  {0.002662(25\pm28)}
\newcommand{\fsrISNmB}{0.00266(42\pm20)}

\newcommand{\DeltaC}{\Delta_\mathrm{C}}

\newcommand{\FF}{\mat{F}}

\newcommand{\zc}{z_\mathrm{c}}
\newcommand{\Kc}{K_\mathrm{c}}
\newcommand{\Tc}{T_\mathrm{c}}
\newcommand{\iso}{\mathrm{iso}}
\newcommand{\hl}{\mathrm{Hl}}
\newcommand{\kB}{k_\mathrm{B}}

\newcommand{\im}{{i{-}1}}

\newcommand{\eulergamma}{\fred{\gamma_\mathrm{EM}}}
\newcommand{\catalan}{G}

\begin{document}

\pagestyle{SPstyle}

\begin{center}{\Large \textbf{\color{scipostdeepblue}{
The anisotropic square lattice Ising model \\ with quenched surface disorder
}}}\end{center}

\begin{center}\textbf{
Luca Cervellera\textsuperscript{1\XX{$\dagger$}},
Oliver Oing\textsuperscript{1},
Jan Büddefeld\textsuperscript{1},
Alexander Hahn\textsuperscript{1} and
Alfred Hucht\textsuperscript{1$\star$}
}\end{center}

\begin{center}
{\bf 1} Fakultät für Physik, Universität Duisburg-Essen and CENIDE, 
D-47048 Duisburg, Germany
\\[\baselineskip]
$\star$ \href{mailto:fred@thp.uni-due.de}{\small fred@thp.uni-due.de}
\end{center}

\section*{\color{scipostdeepblue}{Abstract}}
\textbf{\boldmath%
{
Using exact enumeration, the surface free energy as well as the critical Casimir amplitude and Casimir force are calculated for the square lattice Ising model with anisotropic couplings and quenched surface disorder on one surface in cylinder geometry at criticality.
The system shape is characterized by the aspect ratio $L/M$, where the cylinder length $L$ can take arbitrary values, including $L\to\infty$, while the circumference $M$ is varied from $M=4$ to $M=54$, resulting in up to $2^{54}$ numerically exact free energy calculations. 
A careful $M\to\infty$ extrapolation shows that quenched surface disorder is irrelevant in two dimensions, but gives rise to logarithmic corrections.
\CC{
Ich habe das Gefühl der Abstract ist noch etwas zäh, weil die Sätze so lang sind. Ich habe mal versucht es etwas abzuändern, bin damit aber auch nicht zu 100\% zufrieden. Was meinst du, Fred?
\fred{Sind leider neun Zeilen, s.u. Wir lassen das jetzt erstmal so, ich finde es OK.}
}}
\CC{
We investigate the effect of quenched surface disorder on one of the surfaces in the square lattice Ising model on the cylinder. Using exact enumeration, the surface free energy as well as the critical Casimir amplitude and Casimir force are calculated at criticality.
Our analysis includes anisotropic couplings and multiple types of disorder ensembles. 
The system shape is characterized by the aspect ratio $L/M$, where the cylinder length $L$ can take arbitrary values, including $L\to\infty$.
the circumference $M$ is varied from $M=4$ to $M=54$, resulting in up to $2^{54}$ numerically exact free energy calculations. 
A careful $M\to\infty$ extrapolation shows that quenched surface disorder is irrelevant in two dimensions, but gives rise to logarithmic corrections.
}
}

\vspace{\baselineskip}

\noindent\textcolor{white!90!black}{%
\fbox{\parbox{0.975\linewidth}{%
\textcolor{white!40!black}{\begin{tabular}{lr}%
  \begin{minipage}{0.6\textwidth}%
    {\small Copyright attribution to authors. \newline
    This work is a submission to SciPost Physics. \newline
    License information to appear upon publication. \newline
    Publication information to appear upon publication.}
  \end{minipage} & \begin{minipage}{0.4\textwidth}
    {\small Received Date \newline Accepted Date \newline Published Date}%
  \end{minipage}
\end{tabular}}
}}
}

\XX{


\vspace{10pt}
\noindent\rule{\textwidth}{1pt}
\tableofcontents
\noindent\rule{\textwidth}{1pt}
} 




\section{Introduction}
The square lattice Ising model, first considered 1924 in the doctoral thesis of Ernst Ising \cite{DrIsing,Ising25}, is often referred to as the drosophila of statistical physics. 
It is one of the few exactly solvable models that has a continuous phase transition at a finite temperature.
After the seminal exact solution on the torus by Onsager and Kaufman \cite{Onsager44,Kaufman49}, which lead to the exact bulk free energy density $\fb(T)$ for arbitrary temperature $T$, several other geometries and boundary conditions (BCs) were examined.
Using the dimer representation by Kasteleyn \cite{Kasteleyn61,Kasteleyn63} and its generalization by Fisher \cite{Fisher66}, McCoy and Wu derived an exact solution on the cylinder \cite{McCoyWu73,McCoyWu14}, and gave exact expressions for the surface free energy density $\fs[(o)](T)$ for an open boundary, as well as in a surface magnetic field.
Motivated by the universal critical Casimir effect \cite{FisherdeGennes78}, which describes the attraction or repulsion of boundaries enclosing a medium with correlated fluctuations similar to the quantum electrodynamical Casimir effect \cite{Casimir48}, several other BCs has been studied. For an overview see references \cite{Gambassi09a,DantchevDietrich2023}.

Apart from that, progress was made towards an exact solution of the anisotropic square lattice Ising model without periodic BCs in either direction. 
These calculations turned out to be quite involved, because no Fourier transformation could be used in at least one direction, and were found to be of a similar or even higher complexity compared to the calculation of the bulk magnetization \cite{Yang52}, involving Toeplitz determinants with a size-dependent symbol function \cite{Hucht21a}.
In two completely independent works, Baxter \cite{Baxter16,Baxter20} and Hucht \cite{Hucht16a,Hucht16b,Hucht21a} derived explicit determinantal expressions for the partition function of the Ising model on the rectangle. 
While Baxter used the spinor method by Kaufman and focused on the surface and corner free energy contributions in the thermodynamic limit, verifying a conjecture for the corner contributions by Vernier and Jacobsen \cite{VernierJacobsen12}, Hucht utilized the dimer method combined with block transfer matrices \cite{Molinari08} and focused on the universal Casimir free energy contributions in this system.

The methods established in \cite{Hucht16a,Hucht16b,Hucht21a} turned out to be applicable to other BCs, as well as to disordered systems \cite{MasterBueddefeld,BachelorCervellera,MasterCervellera,BachelorOing}. 
In this work, they will be extended in order to calculate the partition function, the free energy, and the Casimir contributions for a cylinder of arbitrary length $L$ and finite circumference $M\leq54$ with arbitrary fixed BCs at one side, while the other side has open BCs for simplicity. 
By averaging the free energy over all $2^M$ possible boundary configurations, the quenched disorder average is obtained exactly.
Performing a careful extrapolation of these numerically exact results to $M\to\infty$, we determine the random surface free energy and the critical Casimir contributions, and test several predictions for this system \cite{Cardy1991, Diehl97a}.
We remark that our exact approach is orders of magnitude faster than the Monte Carlo simulations performed by Pleimling and coworkers \cite{Pleimling2004}, as we can determine the numerically exact free energy of a system with size $M\times L$ in $\mathcal{O}(M)$ steps using the presented Woodbury tree algorithm, being similar to the principal minor algorithm by Griffin and Tsatsomeros \cite{GriffinTsatsomeros2006}, while the Monte Carlo approach required a full simulation for each boundary configuration.
The principal minor algorithm \cite{GriffinTsatsomeros2006} has also been successfully applied recently to diagrammatic Monte Carlo quantum simulations \cite{SimkovicFerrero2022}.


\section{Model and method}

\subsection{
Quenched surface disorder}

\begin{figure}[t]
	\centering
	\includegraphics[width=0.7\textwidth]{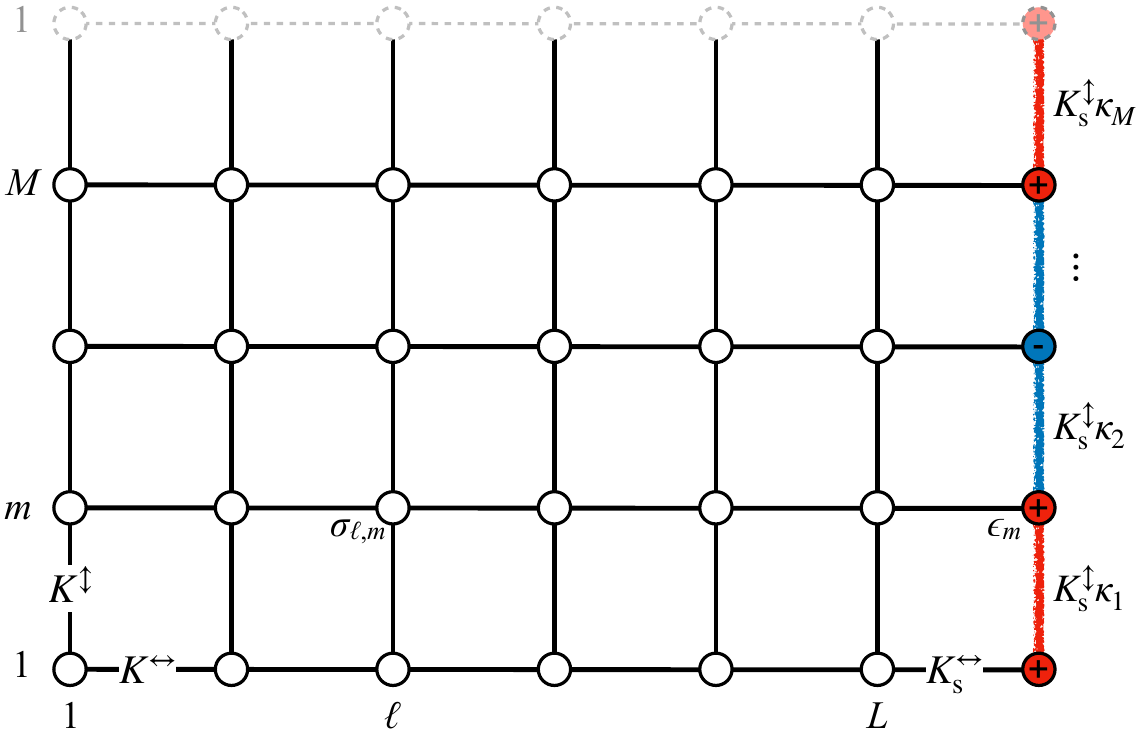}
	\caption{The anisotropic Ising system \eqref{eq:Hok} with $L \times M$ spins $\sigma_{\ell,m}$ on the cylinder, with open boundary conditions on the left side and quenched surface disorder on the right side, shown as fixed red $(+)$ and blue $(-)$ boundary spins $\epsilon_m$.
    The spins $\epsilon_m$ are fixated by a set of infinitely strong couplings $K^\MM_\mathrm s \kappa_{m}$ with sign $\kappa_{m}=\pm1$, see text.
    }
	\label{fig:system}
\end{figure}

The system under consideration is the anisotropic square lattice Ising model on a cylinder with length $L$ and circumference $M$, as sketched in Fig.~\ref{fig:system}.
Neighbouring Ising spins $\sigma_{\ell,m}=\pm 1$ interact via the reduced couplings $K^{\delta} \equiv J^{\delta}/(\kB T)$ in the two directions $\delta = \LL,\MM$, with exchange couplings $J^\delta$, Boltzmann constant $\kB$, and temperature $T$.
While the left side of the cylinder has open boundary conditions (BCs) for simplicity, at the right side we assume quenched randomly disordered BCs, where \XX{\luca{Bcs. The}\fred{Das aendert den Sinn}} the surface spins $\sigma_{\ell,m}$ in layer $L$ couple to frozen fixed boundary spins $\epsilon_m=\pm 1$ in the additional boundary layer $L+1$, with coupling strength $K^\LL_\mathrm s$. 
Note that this boundary layer acts like a quenched random surface field of strength $K^\LL_\mathrm s$.
\XX{fred{Finde ich nicht.}\oo{Den Satz kann man sich glaub ich sparen: Wenn ich das richtig überblicke, wird $K^\LL_\mathrm s$ einmal nach 0 diskutiert (direkt vor \eqref{eq:Zoo_1}) und anschließend explizit auf $K^\LL$ gesetzt (direkt vor \eqref{eq:Zok_2})}}Unless otherwise stated, we assume $K^\LL_\mathrm s=K^\LL$.

We neither can apply symmetry breaking fields nor set certain spins to fixed values in the exact derivation of the partition function.
To overcome this, we add an additional coupling term to the Hamiltonian.
This couples the boundary spins $\epsilon_m$ infinitely strong as required, using the corresponding bond disorder variables $\kappa_{m} \equiv \epsilon_{m} \epsilon_{m+1}$. 
Note that due to the periodic boundary condition in $\MM$ direction the product $\prod_{m=1}^M \kappa_m$ must be one, such that there are $M-1$ independent disorder degrees of freedom $\kappa_1,\ldots\kappa_{M-1}$, and $\kappa_M=\prod_{m=1}^{M-1} \kappa_m$.
In the following we write $\vec\kappa \equiv \{\kappa_1,\ldots,\kappa_{M}\}$ for the disorder configuration.
Using the superscript $\OK$ for the BCs at the two ends of the cylinder, the\XX{ \fred{reduced? oder $\beta H$?}} Hamiltonian is therefore given by
\begin{align}\label{eq:Hok}
    \beta\mathcal{H}^\OK(\vec\kappa)
    =&{}-K^\LL \sum_{\ell=1}^{L-1} \sum_{m=1}^M \sigma_{\ell,m}\sigma_{\ell+1,m}
    -K^\MM \sum_{\ell=1}^L     \sum_{m=1}^M \sigma_{\ell,m}\sigma_{\ell,m+1} \nonumber\\
    &{}-K^\LL_\mathrm s \sum_{m=1}^M \sigma_{L,m}\epsilon_{m}
    -\lim_{K^\MM_\mathrm s\to\infty} K^\MM_\mathrm s \sum_{m=1}^{M} (\kappa_{m} \epsilon_{m}\epsilon_{m+1}-1)\,.
\end{align}
First we discuss two special cases: for $\kappa_m=+1$ all boundary spins $\epsilon_m$ are forced to be parallel, which corresponds to the well known plus ($\epsilon_m=+1$) and minus ($\epsilon_m=-1$) boundary conditions. 
On the other hand, the case $\kappa_m=-1$, which is only allowed for even $M$, resembles the two possible staggered boundary conditions $\epsilon_m=\pm (-1)^m$, which are known to be in the Dirichlet surface universality class \cite{HobrechtHucht18b}.

\subsection{Derivation of the partition function}
\subsubsection{Definitions}

Following the literature \cite{Baxter82} as well as our previous work \cite{Hucht16a, Hucht16b, Hucht21a} we describe the system in terms of the dual couplings
\begin{align}\label{eq:tz_def}
    z &\equiv \tanh{K^\LL} = (\ee^{-2 K^\LL})^* ,&
    t &\equiv (\tanh{K^\MM})^* = \ee^{-2 K^\MM},
\end{align}
and we use the abbreviations \cite{Hucht16a}
\begin{align}\label{eqs:dual_pm}
    a^* &\equiv \frac{1-a}{1+a},
    & 
    a_\pm &\equiv \frac{a\pm a^{-1}}{2},
\end{align}
in the following.
The main focus of this work is on systems at the anisotropic critical point $\zc$, where $z$ and $t$ are equal,
\begin{align}\label{eq:zc_tz_def}
    \zc \equiv t &= z.
\end{align}
Note that $0<\zc<1$ depending on the coupling anisotropy, while the isotropic case $\Kc^\LL=\Kc^\MM$ is given by the critical coupling $\zc^\iso=\sqrt{2}-1$, leading to the famous isotropic critical temperature $\kB \Tc^\iso=2J^\iso/\log(1+\sqrt{2})$ originally derived by Kramers and Wannier \cite{KramersWannier41}. 

Coupling anisotropies are best described geometrically, using the bulk correlation length amplitude ratio \cite{Hucht02a,BrandHucht2023,BrandHucht2024} (we assume lattice constants $a^\LL=a^\MM=1$)
\begin{align}\label{eq:rxi_def}
    r_\xi \equiv \frac{\xi_0^\LL/a^\LL}{\xi_0^\MM/a^\MM}
    = \sinh(2\Kc^\LL) = \frac1{\sinh(2\Kc^\MM)} = \frac{-1}{(\zc)_-} , 
\end{align}
with the anisotropic paramagnetic bulk correlation length\footnote{$f(x)\simeq g(x)$ means \emph{asymptotically equal} in the respective limit, $\lim_{x\to x_0} f(x)/g(x) = 1$.}
\begin{align}\label{eq:xi_def}
    \xi^\delta(\tau) &\stackrel{\tau>0}{\simeq} \xi_0^\delta \tau^{-\nu}&
    \text{in direction } \delta &= \LL,\MM \, ,
\end{align}
reduced temperature $\tau \equiv T/\Tc-1$, and $2d$-Ising correlation length exponent $\nu=1$. The only relevant shape parameter of the considered anisotropic system with given geometric 
aspect ratio $r \equiv (L/a^\LL)/(M/a^\MM)$ is therefore the \emph{reduced aspect ratio} \cite{Hucht02a, HuchtGruenebergSchmidt11, HobrechtHucht18a, HobrechtHucht18b, Hucht16b}
\begin{align}\label{eq:rho_def}
    \rho\equiv\frac{r}{r_\xi} 
    = \frac{L/\xi_0^\LL}{M/\xi_0^\MM} \, ,
\end{align}
which measures the system sizes in terms of the respective correlation length amplitudes $\xi_0^\delta$ from \eqref{eq:xi_def}.
In isotropic systems $r_\xi^\iso \equiv 1$ holds, while in anisotropic systems we can rescale the lattice constant $a^\LL$ according to 
$a^\LL \mapsto \tilde a^\LL = a^\LL/r_\xi$
and find 
isotropic $\tilde r_\xi=1$, while $\tilde\rho=\tilde L/(M \tilde r_\xi)=\rho$ remains invariant, i.\,e., near criticality the coupling anisotropy can be eliminated by a proper rescalation of the system.

In addition to the isotropic critical point, we also consider the so called \emph{Hamiltonian limit} \cite{Henkel1990,Henkel99} where $\zc \to \zc^\hl = 1$. 
In this particular limit, the criticality condition \eqref{eq:zc_tz_def} implies that the $\LL$ couplings go to $\Kc^\LL\to\infty$ and the $\MM$ couplings go to $\Kc^\MM\to0$, while the critical temperature remains finite. 
The correlation length amplitude ratio \eqref{eq:rxi_def} diverges like $r_\xi\simeq(1-\zc)^{-1}\to\infty$ in this limit.
Assuming lattice constants $a^\LL=a^\MM=1$ and requiring constant $M$ and constant $\rho = L/(M r_\xi)$ from \eqref{eq:rho_def}, we find $L=\rho M r_\xi \to\infty$ in this limit, too.
This particular limit can be interpreted as a continuum limit in the $\LL$ direction if we rescale 
the lattice constant $a^\LL=1$ according to 
$a^\LL\mapsto \tilde a^\LL = a^\LL/r_\xi \simeq 1-\zc\to0$.
It turns out in the following that, in the Hamiltonian limit, 
many quantities become particularly simple.


\subsubsection{The block transfer matrix}

The partition function $Z^\OK(L,M;\vec\kappa)$ of the considered system \eqref{eq:Hok} with arbitrary, quenched boundary condition $\vec\kappa$,
as well as the corresponding reduced free energy (in units of $\kB T$),
\begin{align}\label{eq:Fok_1}
    F^\OK(L,M;\vec\kappa)=-\log Z^\OK(L,M;\vec\kappa) \, ,
\end{align}
can be calculated with a transfer-matrix method similar to the case of open BCs on all sides \cite{Hucht16a, Hucht16b, Hucht21a}, with a $2M \times 2M$ block transfer matrix 
\CC{The open and the quenched boundary condition at the two ends of the cylinder are described by the $2M\times M$ block vectors $ |\bmat{e}_{\mathrm{o}} \rangle$ and $ |\bmat{e}_{\kappa} \rangle$, respectively. 
}
%
\begin{align}\label{eq:T_def}
    \bmat{T}=\Ut \Uz \, .
\end{align}
It consists of two $2\times 2$ block matrices with $M\times M$ blocks
\begin{align}\label{eq:Us}
 \Ut = 
 \begin{bmatrix} \mat{H}^- & 0 \\ 0 & \ii\mat{1} \end{bmatrix}^\dagger 
 \begin{bmatrix} \mat{t}_+ & \mat{t}_- \\ \mat{t}_- & \mat{t}_+ \end{bmatrix} 
 \begin{bmatrix} \mat{H}^- & 0 \\ 0 & \ii\mat{1} \end{bmatrix},\quad
  \Uz =
 \begin{bmatrix} \mat{1} & 0 \\ 0 & \ii\mat{1} \end{bmatrix}^\dagger 
 \begin{bmatrix} \mat{z}_+ & -\mat{z}_- \\ -\mat{z}_- & \mat{z}_+ \end{bmatrix} 
 \begin{bmatrix} \mat{1} & 0 \\ 0 & \ii\mat{1} \end{bmatrix},
\end{align}
describing one layer of bulk couplings in $\MM$ and $\LL$ direction, respectively. 
The $M\times M$ diagonal matrices $\mat{t}_\pm=t_\pm \mat{1}$ and $\mat{z}_\pm=z_\pm \mat{1}$ are defined in terms of the couplings $t$ and $z$ \eqref{eq:tz_def} using the $\pm$-notation from \eqref{eqs:dual_pm}, while the skew-circulant shift matrix
\begin{align}
 \mat{H}^- =
 \begin{pmatrix}
  0 & 1                   \\
    & \ddots & \ddots     \\
    &        & \ddots & 1 \\
 -1 &        &        & 0
 \end{pmatrix}. 
\end{align}
accounts for the periodic BCs in $\MM$ direction.
Note that this derivation is also applicable to other BCs in $\MM$ direction by adjusting the lower left element of $\mat H^-$ \cite[Eq.~(7)]{Hucht16a}.
We point out that the block transfer matrices defined above are Wick-rotated w.r.t.\ the definition in \cite{Hucht16a}, resulting in \emph{complex orthogonal} matrices fulfilling $\bmat A^\T$$\bmat A = \bmat 1$, see \cite{Hucht21a} for details.

\CC{\fred{The following must be done later!} In the Hamiltonian limit $\zc\to1$, even for finite systems the free energy contribution of the adapted constant $C_2 = C_2^\dag/(2 \z_{\mathrm{c}-}^M) = 2^{M-1} (-2/{\zc}_-)^{LM}$, with $C_2^\dag$ from \cite[Eq.(25c)]{Hucht16a}, diverges \fred{Luca:check!}. In order to get continuous and finite results in this limit, we replace $C_2$ by $C = C_2 r_\xi^{-LM} = 2^{M-1} 2^{LM}$ in the following, such that $C=C_2^\iso$ is unchanged in the isotropic case.}\XX{\fred{FIXHL}}
With that, the partition function of the cylinder with open BCs on both sides, i.e., for $K^\LL_\mathrm s\to0$ and dropped last term in \eqref{eq:Hok}, is given by
\begin{align}
    \label{eq:Zoo_1}
    \bigl[Z^\OO(L,M)\bigr]^2 = 
    2^{2M}\left(\frac{-2}{z_-}\right)^{(L-1)M}
    \det\langle\bmat{e}_\mathrm o |  \bmat{T}^{L-1} \Ut |\bmat{e}_\mathrm o \rangle\,,
\end{align}
where 
the $2M\times M$ boundary block vector
\begin{align}
\label{eq:eo_def}
 |\bmat{e}_{\mathrm{o}}\rangle &= \frac{1}{\sqrt{2}}
\begin{bmatrix}
\mat{1} \\
 -\ii\mat{1}
\end{bmatrix},& 
 \langle\bmat{e}_{\mathrm{o}} |\bmat{e}_{\mathrm{o}}\rangle &= \mat{1} \, ,
\end{align}
describes the open surface. 
We use Bra-ket notation, where $\langle\,\cdot\,|$ denotes the conjugate transpose of $|\,\cdot\,\rangle$, and \eqref{eq:Zoo_1} is correct for both even and odd $M$ and $L$.

We now turn to the system \eqref{eq:Hok}, with boundary disorder $\vec\kappa$. For simplicity we assume $K^\LL_\mathrm s=K^\LL$ in the following. 
The partition function can thus be written as
\begin{align}
    \label{eq:Zok_2}
    \bigl[Z^\OK(L,M;\vec\kappa)\bigr]^2 =
    2^{2M} \left(\frac{-2}{z_-}\right)^{LM}
     \det\langle\bmat{e}_\mathrm o | \bmat{T}^{L} t_\mathrm s \Ut_\kappa |\bmat{e}_\mathrm o \rangle\,,
\end{align}
and the trace was performed over both $\sigma_{\ell,m}$ and $\epsilon_m$.
Here, we defined the matrix $\Ut_\kappa$ according to \eqref{eq:Us}, but with the replacement $\mat t_-\mapsto \mat\kappa \mat t_{\mathrm s -}$, where $\mat\kappa = \diag(\vec \kappa)$ and $t_\mathrm s = \ee^{-2K^\MM_\mathrm s}$.

In order to obtain a system with a quenched boundary condition on the right side, the boundary couplings $K^\MM_{\mathrm s,m}=\kappa_m K^\MM_{\mathrm s}$ have to be infinitely strong, where each coupling sign $\kappa_m=\pm 1$ can later be chosen independently to create arbitrary boundary conditions.
The corresponding limit $\lim_{t_\mathrm{s} \to 0} \Ut_\kappa $ is regularized by the factor $t_\mathrm s$ 
in order to get a finite result,
\begin{align}
\label{eq:e_kappa}
    \lim_{t_\mathrm{s} \to 0} t_\mathrm s\Ut_\kappa &= |\bmat{e}_\kappa\rangle\langle\bmat{e}_\kappa|\, ,& 
     \text{with} \quad
    |\bmat{e}_\kappa\rangle &= \frac{1}{\sqrt{2}}
\begin{bmatrix}
\mat{1} \\
 \ii\bm{\kappa} \mat{H}^-
\end{bmatrix}, &
 \langle\bmat{e}_{\kappa}  |\bmat{e}_{\kappa}\rangle &= \mat{1} \, .
\end{align}
The resulting block vector $|\bmat{e}_\kappa\rangle$ describes an arbitrary quenched boundary condition. 
Inserting \eqref{eq:e_kappa} into \eqref{eq:Zok_2} results in the partition function for an arbitrary quenched boundary%
\bS\begin{align}
    \bigl[Z^\OK(L,M;\vec\kappa)\bigr]^2
    &= 2^{2M} \left(\frac{-2}{z_-}\right)^{LM}
    \det\langle\bmat{e}_\mathrm o| \bmat{T}^L |\bmat{e}_\mathrm \kappa \rangle \, \det\langle\bmat{e}_\mathrm \kappa  |\bmat{e}_\mathrm o\rangle\\
    &= C\det\langle\bmat{e}_\mathrm o |  \bmat{T}^L |\bmat{e}_\mathrm \kappa \rangle
     \, , \label{eq:Z(o,r)_2}
\end{align}\eS%
with constant $C=2^{M+\XX{\luca{-}}1} (-2/z_-)^{LM}$.\XX{\marginpar{\luca{-ich habs nochmal gecheckt, bei mir kommt da ein - hin.}}}
Note that in the Hamiltonian limit, $C$ needs to be regularized as discussed in  Section~\ref{sec:Hl}.
We used $\det\langle\bmat{e}_\mathrm \kappa|\bmat{e}_\mathrm o\rangle =  2^{1-M}$, 
as the product over all $\kappa_m$ is always one by definition. 
Due to the quenching, the $M$ previously free boundary spins $\epsilon_m$ become the arbitrary fixed boundary, which results in a reduction of the partition function by a factor of $2^{M/2}$.
Furthermore, any boundary condition characterized by coupling signs $\kappa_m$ represents two $Z_2-$symmetric spin configurations, resulting in an additional factor of $2$ in the partition function to get to \eqref{eq:Z(o,r)_2}.

\subsubsection{The Onsager dispersion relation}

In the following, we use the usual \cite{Baxter82,Hucht16a} parametrisation of the model through the real angles $\varphi$ and $\gamma$, which are intimately related through the famous Onsager dispersion relation of the anisotropic square lattice Ising model,
\begin{align}\label{eq:Onsager}
    \cosh\gamma+t_-z_-\cos\varphi=t_+z_+
    \, . 
\end{align}
The hyperbolic angle $\gamma$ is linked to the eigenvalues $\lambda_{\bmat{T},\mu}=\ee^{\gamma_\mu}$ of the transfer matrix $\bmat{T}$,
while the trigonometric angle $\varphi$ is connected to the eigenvalues of the transfer matrix propagating in $\MM$ direction, see, e.g., Ref.~\cite[Eq.~(41)]{Hucht16a}.
Due to the periodic boundary conditions in $\MM$ direction, the angle $\varphi$ takes the simple form 
\begin{align}\label{eq:phi_mu}
    \varphi_\mu = \frac{\pi \mu}{M},
\end{align}
where from now on $\mu$ runs over the $M$ odd integers 
$\mu\in\{1,3,5,\ldots,2M-1\}$, such that $0<\varphi_\mu<2\pi$.
The following derivation is valid for both even and odd $M$. If $M$ is odd, the special case $\varphi_M = \pi$ must be taken care of.
We refrain from using the representation through elliptic function as derived in \cite{Hucht21a}, as we only discuss the anisotropic critical point here.

\subsubsection{The angles $\alpha$ and $\psi$}

From now on we focus on the anisotropic critical point $\zc\equiv z=t$ from \eqref{eq:zc_tz_def}.
As shown below, a very simple form for the quantities of interest at criticality can be obtained by defining the
angle $\alpha$ via the relation
\begin{align}\label{eq:alpha}
    \cos\frac \alpha 2 = 
    \frac{1-\zc^2}{1+\zc^2} \cos\frac\varphi 2.
\end{align}
The angle $0<\alpha<2\pi$ encodes the anisotropy of the critical point, and $\alpha=\pi$ in the Hamiltonian limit $\zc\to1$, cf.\ section \ref{sec:Hl}.
With this definition, the Onsager dispersion relation \eqref{eq:Onsager} at criticality becomes%
\bS\label{eqs:Onsager_c}\begin{align}
    \cosh \gamma &=
     (\zc)_+^2 - (\zc)_-^2 \cos\varphi\\
    &=
    \left[\frac
    {\sin\frac{\alpha+\varphi}{2}}
    {\sin\frac{\alpha-\varphi}{2}}
    \right]_+
    =\frac 1 2
    \left(
    \frac
    {\sin\frac{\alpha+\varphi}{2}}
    {\sin\frac{\alpha-\varphi}{2}}
    +\frac
    {\sin\frac{\alpha-\varphi}{2}}
    {\sin\frac{\alpha+\varphi}{2}}
    \right) \,,
\end{align}\eS%
such that the Onsager-$\gamma$ obey the simple form
\begin{align}
    \ee^\gamma &=\frac
    {\sin\frac{\alpha+\varphi}{2}}
    {\sin\frac{\alpha-\varphi}{2}} \, .
    \label{eq:gamma}
\end{align}
These relations will be derived in the next section.
We finally introduce the $L$-dependent angle $\psi\in\mathbb{R}$ via the relation
\begin{align}\label{eq:psi}
    \tan\frac\psi 2
    &= \frac{\sin\frac\varphi2 + \sin\frac\alpha 2 \coth(L \gamma)}{\cos\frac\varphi2 - \cos\frac\alpha 2} \, ,
\end{align}
that will considerably simplify the expressions below.

\subsubsection{Diagonalization}

The diagonalization of the block transfer matrix $\bmat T$ is done in two steps. 
Due to the periodic BCs in $\MM$ direction we can diagonalise the four blocks simultaneously 
using the discrete Fourier transform
\begin{align}
    \label{eq:fourier}
     \FF=\frac{1}{\sqrt{M}}\Big[ \, \ee^{\ii m \varphi_\mu}\Big]_{m=0,\; \mu=1}^{M-1,\;2M-1} \, ,
\end{align}
with angle $\varphi$ from \eqref{eq:phi_mu}. 
We apply the block diagonal matrix
\newcommand{\sfp}{{\color{red}+}}
\newcommand{\sfm}{{\color{red}-}}
\begin{align}
\bmat F = \begin{bmatrix}
\mat F & \mat 0\\
\mat 0 & \mat F
\end{bmatrix}
\end{align}
to $\bmat T$ in order to diagonalise the blocks, getting%
\bS\begin{align}\bmat t_\mu &\equiv 
(\bmat F^\dag \bmat T \bmat F)_{\mu\mu} =
\begin{bmatrix}
       \tau_{\mu}^+      & -\ii \tau_{\mu}^-      \\
    \ii\tau_{\mu}^{-\,*} &      \tau_{\mu}^{+\,*}
\end{bmatrix}, 
\intertext{with}
    \tau_{\mu}^\pm &= t_+ z_\pm - t_- z_\mp \ee^{-\ii\varphi_\mu}\,.
\end{align}\eS%
From now on we focus on the anisotropic critical point and replace $t$ and $z$ by $\zc$, which can vary from zero to one. The generalisation to non-critical temperatures is straightforward.
At criticality we can diagonalise the $2\times2$ block matrix by applying the matrix of eigenvectors $\bmat x_\mu$ and the rotation matrix $\bmat r_{\phi}$,
\begin{align}
    \bmat x_\mu &=
    (\bmat X)_{\mu\mu} = \frac{\ee^{-\frac\ii4\varphi}}{\sqrt{2\sin\frac\alpha2}}
    \begin{bmatrix}
        \ee^{ \frac{\ii}{2}\alpha_\mu} &
        -\ee^{-\frac{\ii}{2}\alpha_\mu} \\
        \ii\ee^{ \frac{\ii}{2}\varphi_\mu} &
        -\ii\ee^{ \frac{\ii}{2}\varphi_\mu}
    \end{bmatrix} , 
    &
    \bmat r_{\phi} &= \begin{bmatrix}\cos\phi&\sin\phi\\-\sin\phi&\cos\phi\end{bmatrix},
\end{align}
with $\alpha$ from \eqref{eq:alpha}, fulfilling $\bmat x_\mu^{-1}= \bmat r_{\pi/2}^\T \bmat x_\mu^\T \bmat r_{\pi/2}$,
to finally get the eigenvalues of the block transfer matrix $\bmat T$,
\begin{align}
\lambda_{\bmat T,\mu} = (\bmat X^{-1}\bmat F^\dag \bmat T \bmat F \bmat X)_{\mu\mu} =
\begin{bmatrix}
 \ee^{-\gamma_\mu} & 0 \\
 0 & \ee^{+\gamma_\mu}
\end{bmatrix},
\end{align}
with $\gamma$ from \eqref{eq:gamma}.
In conclusion, the real valued angles $\gamma$ and $\alpha$ naturally appear in the eigenvalues and eigenvectors of the critical transfer matrix $\bmat T$.
With these results we now can derive%
\bS\begin{align}
   \bmat{T}^L &=
   \begin{bmatrix}
      \mat P_1 & \ii \mat P_2^\T \\
 -\ii \mat P_2 &     \mat P_1^\T
   \end{bmatrix}
   ,
\end{align}
with real $M\times M$ matrices $\mat P_{1,2}$ having the eigenvalues
\begin{align}
    \lambda_{\mat P_1} &= \ii \frac{\sinh\big(L\gamma-\tfrac\ii2 \alpha\big)}{\sin\tfrac\alpha2},
    &
    \lambda_{\mat P_2} &= -\ii  \ee^{\frac\ii2\varphi} \frac{\sinh\big(L\gamma\big)}{\sin\tfrac\alpha2}.
\end{align}\eS%
From now on we are analytic in $L\in\mathbb R^+$.

\subsubsection{Factorisation of the partition function}

Defining the 
matrix $\mat P = \frac{1}{2}(\mat P_1 + \mat P_2)$, with angle $\psi$ from \eqref{eq:psi} and with eigenvalues \XX{\fred{fix}}%
\bS\begin{align}
    \lambda_{\mat P,\mu}
    = \left( \FF^\dag \mat P \FF \right)_{\mu,\mu} 
    \label{eq:lambdaP_1} 
    &=
    \frac{\ee^{\frac\ii2\varphi} \, \sinh\big(L\gamma\big) - \sinh\big(L\gamma-\tfrac\ii2 \alpha\big)}{2\ii\sin\tfrac\alpha2} 
    \\ \label{eq:lambdaP_2}
    &= \frac{\sin\frac\alpha2 \cosh(L\gamma) + \sin\frac\varphi2 \sinh(L\gamma)}{\sin\frac\alpha2\, (1-\ee^{-\ii\psi})},
\end{align}\eS%
and evaluating the matrix product $\langle\bmat{e}_\mathrm o |  \bmat{T}^L |\bmat{e}_\mathrm \kappa \rangle=\mat P - \mat P^\T \mat\kappa \mat H^-$, the partition function \eqref{eq:Z(o,r)_2} simplifies to
\CC{
OLD: Now, further focus will be laid on rewriting \eqref{eq:Z(o,r)_2} to handle the exponential growth of the elements of $\bmat{T}^L$ in $L$. 
Its block matrix representation reads
\begin{align}
   \bmat{T}^L=
   \begin{bmatrix}
\mat P_1& \ii \mat P_2\\
 -\ii \mat P_2^\T & \mat P_1^\T
\end{bmatrix},
\end{align}
with real $M\times M$ matrices $\mat P_i$ that can be calculated analytically from \eqref{eq:Us}. 
Defining the $M\times M$ matrix $\mat P = \frac{1}{2}(\mat P_1 + \mat P_2^\T)$, the partition function \eqref{eq:Z(o,r)_2} simplifies to%
}
\bS\begin{align}
    \bigl[Z^\OK(L,M;\vec\kappa)\bigr]^2 
    &= C \det\left(\mat P - \mat P^\T \mat\kappa \mat H^-\right) \\
    &= C \det \mat P \, \det\left(\mat 1 - \mat P^{-1} \mat P^\T \mat\kappa \mat H^-\right)\\
    &= C \det \mat P \, \det\left(\mat 1 - \mat H^- \mat P^{-1} \mat P^\T \mat\kappa \right)\\
    \label{eq:Zor}
    &= C \det \mat P \, \det\left(\mat Q + \mat\kappa \right) \, .
\end{align}\eS%
While $\mat P$ grows exponentially in $L$, the matrix elements of
\begin{align}
\label{eq:Q_0}
    \mat Q =-\mat H^- \mat P^{-1} \mat P^\T
\end{align}
are $O(1)$ for arbitrary $L$,  $M$ and $\zc$.
As the determinant of $\mat P$ does not depend on $\kappa$, we can divide the partition function by a reference partition function, in order to cancel out the corresponding determinant. 

\subsubsection{The excess free energy}

Hence we define the (reduced) \emph{excess free energy} relative to the staggered BC $\mathrm{(\st)}$ having $\kappa_m=-1$,%
\bS\label{eqs:Fexk}
\begin{align}
F^{\EX[,\kappa]}(L,M;\vec\kappa) &\equiv F^\OK(L,M;\vec\kappa) - F^{\mathrm{(o,\st)}}(L,M) \\
&= -\frac{1}{2}\log \det\left(\mat{Q}+\mat{\kappa}\right)+\frac{1}{2}\log \det\left(\mat{Q}-\mat{1}\right).
\end{align}\eS%
Note that the staggered BC is only well-defined for even $M$, so odd $M$ must be related to, e.g., the $(+)$ BC. We focus on even $M$ in the following.


\CC{
As shown in Appendix \ref{Appendix:partition-function}, we can factorize the determinant in the partition function \eqref{eq:Zok_1} according to 
\begin{align}\label{eq:det_eo_ek}
    \det \langle\mat e_\mathrm o | \bmat{T}^L |\mat e_{\kappa} \rangle
    = \det\mat P \, \det\left(\mat Q + \mat\kappa \right),
\end{align}
where the matrix $\mat P$ contains the contributions from the bulk and the open boundary, $\mat Q$ is a skew-circulant matrix convergent in the limit $L,M\to\infty$, and $\mat\kappa=\diag(\vec\kappa)$ contains the disorder. 
}
\XX{\fred{fix subequations}}
\subsubsection{The matrix $\mat Q$}

The matrix $\mat Q$ given in \eqref{eq:Q_0} is an important quantity in our analysis, as it relates the disorder dependent part of the partition function \eqref{eq:Zor} to the concrete disorder configuration given by the diagonal matrix $\mat\kappa$ through the determinant of $\mat Q+\mat\kappa$.
The skew-circulant matrix $\mat Q$ is convergent in the limit $L\to\infty$ for arbitrary $M$, as well as in the Hamiltonian limit.
The eigenvalues of $\mat Q$ are easily calculated from \eqref{eq:lambdaP_2} and \eqref{eq:Q_0} to be%
\bS\label{eqs:lambda_Q}\begin{align}
    \lambda_{\mat Q,\mu}
    &= \left( \FF^\dag \mat Q \FF \right)_{\mu,\mu} 
    = -\ee^{\ii\varphi} \lambda_{\mat P,\mu}^{-1} \lambda_{\mat P,\mu}^* 
    = -\ee^{\ii \varphi} \frac{1-\ee^{-\ii\psi}}{1-\ee^{+\ii\psi}}\\
    &= \ee^{\ii(\varphi-\psi)}.
\end{align}\eS%
The matrix elements of $\mat Q$ therefore read%
\bS\begin{align}
    (\mat Q)_{ij} &= \frac{1}{M} \sum_{\substack{0<\mu<2M\\\mu\text{ odd}}}\ee^{\ii \left[(i-j) \varphi-\psi \right]}\\
    &= \frac{1}{M} \sum_{\substack{0<\mu\leq M\\\mu\text{ odd}}}(2-\delta_{\mu,M})\cos\left[ (i-j) \varphi-\psi \right] \, ,
\end{align}\eS%
where the Kronecker-$\delta$ ensures that for odd $M$ the eigenvalue at $\mu=M$ is counted correctly.

Finally, we turn to the infinitely long cylinder. 
In the limit $\rho\to\infty$ the eigenvalues \eqref{eqs:lambda_Q} simplify to
\begin{align}
    \lambda_{\mat Q_\infty,\mu} = - \ee^{\tfrac \ii 2 (\alpha+\varphi)},
\end{align}
such that now
\begin{align}
   (\mat Q_{\infty})_{ij} &\stackrel{\hphantom{\zc\to1}}{=} 
   -\frac{1}{M} \sum_{\substack{0<\mu<2M\\\mu\text{ odd}}}\ee^{\ii \left[(i-j) \varphi+\frac {\alpha+\varphi} 2 \right]}\\
    & \stackrel{\zc\to1}{=} \frac{1}{M \sin\big[\frac{\pi}{M}(i-j+\tfrac 1 2)\big]}.
\end{align}

\subsubsection{The Hamiltonian limit}
\label{sec:Hl}

\XX{$L=\rho M r_\xi \to\infty$}

In the Hamiltonian limit $\zc\to1$, where the correlation length amplitude ratio \eqref{eq:rxi_def} $r_\xi\to\infty$, the free energy contribution of the adapted constant $C_2 = C_2^\dag/(2z_{\mathrm{c}-}^M) = 2^{M-1} (-2/{\zc}_-)^{LM}$, with $C_2^\dag$ from \cite[Eq.(25c)]{Hucht16a}, diverges even for finite systems.\XX{\fred{Lass uns das nach dem Urlaub diskutieren.}\luca{Also $2^{LM}$ divergiert an sich und $\det P$ auch. Wenn wir $F^{\EX}$ berechnen kürzt sich $\det P$ weg und der Vorfaktor steht vor $\det(\mat Q+\mat \kappa)$. Da $\mat Q$ nicht mit $2^{-L M}$ nach null geht, müssen wir C=1 setzen, oder wo hab ich da einen Denkfehler?}}\XX{ \fred{Luca:check!} 
\luca{$2^{LM}$ müsste doch schon divergieren oder nicht? Weil $L=(-1/z_-)\rho M$. Ich würde sagen man muss den Faktor im Grenzfall komplett weglassen, da das system durch $z\to1$ kontinuierlich wird.}
}
In order to get continuous and finite results in this limit, we replace $C$ from \eqref{eq:Z(o,r)_2} by $C^\hl = C r_\xi^{-LM} = 2^{M\fred{+}1} 2^{LM}$ in the following, such that $C=C_2^\iso$ is unchanged in the isotropic case, where $r_\xi=1$.
In this limit, the angles $\alpha$, $\gamma$, and $\psi$ defined above become
\begin{align}
\alpha         &\stackrel{\zc\to1}{=} \pi \\
L \gamma       &\stackrel{\zc\to1}{=} 2 M \rho \sin\frac\varphi2 \\
\tan\frac\psi2 &\stackrel{\zc\to1}{=} \frac{\sin\frac\varphi2 + \coth(2M\rho \sin\frac\varphi2)}{\cos\frac\varphi2} \, .
\end{align}

\XX{\fred{Q Formeln}}

Note that we use the modified constant\XX{ \fred{check}} $C = 2^{M-1} 2^{LM}$ thorough this work in order to get continuous and finite results in the Hamiltonian limit, where the usual free energy density would diverge. 
Thereby, we have dropped a factor of $r_\xi^{LM}$ in $C$ with respect to the definition in \cite[Eq.(25c)]{Hucht16a}, which has no effect in the isotropic case where $r_\xi^\iso=1$.

\subsubsection{Force derivative}
For the Casimir force we need the derivative of the free energy w.r.t.~$L$.
The required derivative $\partial\lambda_{\mat Q}/\partial L$ of \eqref{eqs:lambda_Q} can most easily be calculated using the chain rule,
\begin{align}
    \frac{\partial\lambda_{\mat Q}}{\partial L}
    =
    \frac{\partial\lambda_{\mat Q}}{\partial \psi}
    \frac{\partial\psi}{\partial\tan\frac\psi2} 
    \frac{\partial\tan\frac\psi2}{\partial L}\,,
\end{align}
such that
\begin{align}
    \frac{L}{\lambda_{\mat Q}}\frac{\partial \lambda_{\mat Q}}{\partial L}
    = \frac{2\cos^2 \frac\psi2}{\sinh^2(L\gamma)} \, \frac{\ii L\gamma \sin\frac\alpha2}{\cos\frac\varphi2 - \cos\frac\alpha2}.
\end{align}




\subsection{Disorder ensembles and the DOTS}

In the presence of quenched disorder, the reduced excess free energy \eqref{eqs:Fexk}
must be averaged over the disorder ensemble $\{\kappa\}$ according to 
\begin{align}\label{eq:Fbar}
    \bar F^{\EX[,\kappa]}(L,M) = \bigl\langle F^{\EX[,\kappa]}(L,M;\vec\kappa)\bigr\rangle_\kappa
    = \frac{\sum_{\kappa} F^{\EX[,\kappa]}(L,M;\vec\kappa)}{\sum_{\kappa} 1}
\end{align}
in order to get the proper disorder-averaged free energy.  
In this work, we will distinguish different disorder ensembles, discriminated by the \emph{boundary magnetization} (density)
\begin{align}\label{eq:mB}
    \mB(\vec\kappa) \equiv \frac{\MB(\vec\kappa)}{M} = \frac 1 M \sum_{m=1}^M \epsilon_m = \langle \epsilon_m \rangle
\end{align}
of the random boundary: (i) the free ensemble average 
$\langle\,\cdot\,\rangle_\kappa^\mathrm{(r)}$
runs over all $2^{M}$ possible disorder configurations%
, such that $-1 \leq \mB(\vec\kappa) \leq 1$ takes all possible values, while (ii) the ensemble $\langle\,\cdot\,\rangle_\kappa^{(\mB=0)}$ is restricted to the $\frac12 \binom{M}{M/2}$ configurations with fixed $\mB(\vec\kappa)=0$.
Note that in both cases $\mB[\bar]=0$. 
Furthermore, (iii) ensembles with a boundary magnetization fixed  to other values than zero, $\mB(\vec\kappa)=\mB\neq0$, are also considered and denoted accordingly.

Alternatively to the simple average in \eqref{eq:Fbar}, we can describe the distribution of the free energies in more detail by introducing the \emph{density of thermodynamic states} (DOTS),
\begin{align}\label{eq:dots^ex}
    \omega^{\EX}(L,M;F^{\EX},\MB) &\equiv \bigl\langle \delta\bigl[ F^{\EX[,\kappa]}(L,M;\vec\kappa) - F^{\EX} \bigr] \,\delta\bigl[ \MB(\vec\kappa)-\MB \bigr] \bigr\rangle_\kappa,
\end{align}
with Dirac's delta distribution $\delta(\,\cdot\,)$, which measures the number of quenched realizations with given excess free energy $F^{\EX}$ and boundary magnetization $\MB$.
Note the normalization $\iint\! \dd F \,\dd \MB \, \omega^{\EX}(F,\MB)=1$.
We recover the ensemble averages (i)-(iii) defined above as the first moments%
\bS\begin{align}\label{eq:Fbar^r}
    \bar F^{\EX[,r]}(L,M) &= \iint\! \dd F \,\dd \MB \, F \, \omega^{\EX}(L,M;F,\MB) \, ,\\
    \bar F^{\EX[,\mathit{\mB}]}(L,M) 
    &= \frac{\int\! \dd F  \, F \, \omega^{\EX}(L,M;F,\MB)}{\int\! \dd F \, \omega^{\EX}(L,M;F,\MB)} 
    = \frac{2^M}{\binom{M}{\frac{M}{2}(\mB+1)}} \int\! \dd F  \, F \, \omega^{\EX}(L,M;F,\MB) \, .
    \label{eq:Fbar^mB}
\end{align}\eS%

\begin{figure}[t!]
	\centering
	\includegraphics[width=0.75\textwidth]{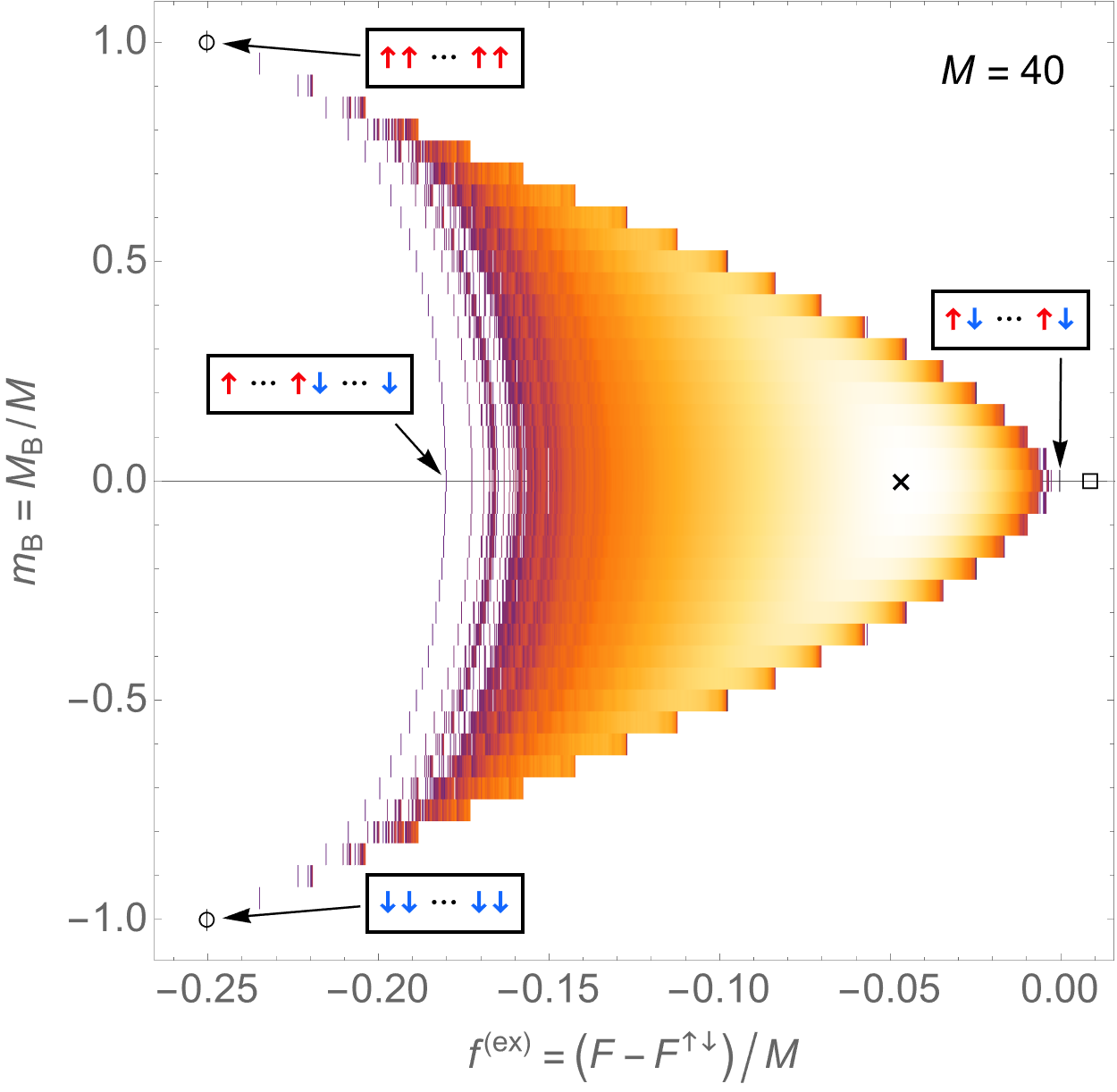}
	\caption{Density of thermodynamic states \eqref{eq:dots^ex}, with excess free energy density $f^{\EX[,\kappa]}$ and boundary magnetisation density $\mB$, for a lattice with $M=40$, $L\to\infty$ and critical isotropic couplings $\zc^\iso$. 
	Exact calculations were done for all $2^M$ surface disorder configurations, with some distinguished boundary configurations marked in the figure.
 Also shown are the surface free energies $\fsp(\zc^\iso)$, $\fsst(\zc^\iso)$ and $\fsr(\zc^\iso)$ from 
 Table \ref{tab:fs} as ($\circ$), ($\scriptstyle\square$) and ($\scriptstyle\times$), respectively.
 }
	\label{fig:DOTS^ex}
\end{figure}

The resulting distribution function for an isotropic system with $M=40$ and $L\to\infty$, containing as much as $2^{40}$ exact free energies, is shown in Fig.~\ref{fig:DOTS^ex}. 
To get a little intuition for this density of thermodynamic states, some distinguished boundary spin configurations $\{\epsilon_m\}$ are marked in the figure.
From Fig.~\ref{fig:DOTS^ex} we can read off the two extremal cases,%
\bS\begin{align}
    \max_\kappa F^{\EX[,\kappa]}(L,M;\vec\kappa) &=F^{\EX[,\st]}(L,M) = 0,\\ 
    \min_\kappa F^{\EX[,\kappa]}(L,M;\vec\kappa) &=F^{\EX[,+]}(L,M),
\end{align}\eS%
such that $F^{\EX[,\kappa]}(L,M;\vec\kappa)\leq 0$ for all $\vec\kappa$.

\subsection{Residual free energy}

In the thermodynamic limit $L,M \rightarrow\infty$, the averaged total free energy \eqref{eq:Fok_1} 
diverges and needs to be regularized to get the finite \emph{residual free energy} \cite{HuchtGruenebergSchmidt11,Hucht16b}
\begin{align}\label{eq:dFbar}
\delta \bar F^\OK(L,M) \equiv \bar F^\OK(L,M) - \bar F^\OK_\infty(L,M)\, .
\end{align}
The leading divergence $\bar F_\infty$ contains the bulk and surface free energies of the system,
\begin{align}\label{eq:Fbar_inf}
\bar F^\OK_\infty(L,M) \equiv L M \fb^{\vphantom{()}} + M(\fs[(o)] + \fsr[(\kappa)])\, ,
\end{align}
where the quenched random boundary must be disorder averaged in order to get a well defined limit for the \emph{random surface free energy density},
\bS\begin{align}\label{eq:fsr}
    \fsr[(\kappa)] \equiv - \fs[(o)] + \lim_{L,M\to\infty} \frac 1 M \big[\bar F^\OK(L,M) - L M \fb \big]\, .
\end{align}
Using the excess free energy \eqref{eq:Fbar} from the last section, we alternatively can write the simpler and numerically more favourable form
\begin{align}\label{eq:fsr2}
    \fsr[(\kappa)] = 
    \fs[(\pm)] + \lim_{L,M\to\infty} \frac 1 M \bar F^{\EX[,\kappa]}(L,M).
\end{align}\eS%
Note that all free energy densities depend on the coupling anisotropy, parametrised through the critical coupling $\zc$ from \eqref{eq:zc_tz_def} in this work.

\begin{figure}
	\centering
	\includegraphics[width=0.8\textwidth]{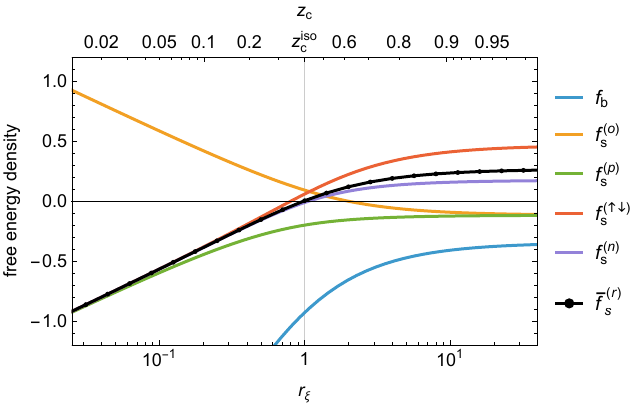}
    \caption{Critical free energy densities from Eqs.~\eqref{eqs:fs_tab} as functions 
    of the near-critical correlation length amplitude ratio $r_\xi$ (Eq.~\eqref{eq:rxi_def}, lower scale) as well as of the critical coupling $\zc$ (upper scale).
    The critical random surface free energy $\fsr(\zc)$ (black) is calculated in Sec.~\ref{sec:fsr}.
    Note that the value at $\zc^\iso$ significantly deviates from zero.
    \label{fig:fs}
    }
\end{figure}

The bulk free energy density $\fb$ and the surface free energy densities $\fs[(\cdot)]$ for open, staggered and fixed surfaces are known exactly both at and away from criticality at least in the isotropic case \cite{Baxter82,McCoyWu14,HobrechtHucht18a,HobrechtHucht18b}.
The exact expressions for all relevant critical free energy densities for arbitrary  anisotropic critical couplings $\zc$ are derived in section~\ref{sec:fs} and read
\bS\label{eqs:fs_tab}
\begin{align}
    \fb(\zc) &= 
    -\frac{1}{4\pi}\int_0^\pi\log\left[
    4 \, \frac{\sin^2 \frac{\alpha+\varphi}{2}}{\sin^2 \frac{\alpha-\varphi}{2}} \right]\dd\varphi \, ,\\
    \fs[(o)](\zc) &=
    -\frac{1}{4\pi}\int_0^\pi\log\left[
    \frac{\tan^2 \frac{\alpha+\varphi}{4}}{\tan^2 \frac{\alpha}{2}}
    \right]\dd\varphi \, ,\\
    \fsst(\zc) &= \fs[(o)](\zc)
    -\frac{1}{4\pi}\int_0^\pi\log\left[
    4\frac{\cos^4 \frac{\alpha+\varphi}{4}}{\cos^2 \frac{\alpha}{2}}
    \right]\dd\varphi \, ,\\
    \fsp(\zc) &= \fsst(\zc)
    -\frac{1}{4\pi}\int_0^\pi\log\left[
    \tan^2 \tfrac{\alpha+\varphi}{4}
    \right]\dd\varphi \, .
\end{align}\eS%
They are shown in Fig.~\ref{fig:fs} as functions of both the critical coupling $\zc$ as well as the correlation length amplitude ratio $r_\xi$ from \eqref{eq:rxi_def}.
The results for the isotropic case as well as for the Hamiltonian limit are given in Table \ref{tab:fs} for reference.
The listed random surface free energy densities \eqref{eq:fsr} will be calculated numerically for different disorder ensembles in section~\ref{sec:fsr}. 
\begin{table}[b]
\centering
\caption{Critical surface free energy contributions for the isotropic system and for the Hamiltonian limit derived from \eqref{eqs:fs_tab}, with Catalan's constant $\catalan$ and 
the Dirichlet $L$ function.
\label{tab:fs}
}
\begingroup
\renewcommand*{\arraystretch}{1.3}
\begin{tabular}{l c|c|c} 
\multicolumn{2}{c|}{Quantity} & 
$\zc^\iso$ & $\zc\to1$ \\
\hline\hline
     bulk & $\fb$ & 
        $-\frac{2\catalan}{\pi}-\frac{\log2}{2}$ & $-\frac{\log2}{2}$\\
\hline
open BC & $\fs[(o)  ]$ & 
    $-\frac{1}{8} L^{(0,0,1)}(8,4,-1)-\frac{3}{8}\log \zc^\iso$ & 
    $-\frac{\catalan}{\pi}+\frac{\log2}{4}$\\
plus BC & $\fs[(+)  ]$ & 
    $\fs[(o)]-\frac{\catalan}{\pi}$ & 
    $-\frac{\catalan}{\pi}+\frac{\log2}{4}$\\
staggered BC & $\fs[(\st)]$ & 
    $-\fs[(o)]-\frac{\catalan}{\pi}-\frac{1}{2}\log \zc^\iso$ 
    & $+\frac{\catalan}{\pi}+\frac{\log2}{4}$\\
internal BC & $\fs[(n)  ]$ & 
   $-\frac{\catalan}{2\pi}-\frac{1}{4}\log\zc^\iso - \frac{\log2}{8}$ 
 & $+\frac{\log2}{4}$\\
 random BC & $\fsr$          & $\fsrISN$   & $\fsrHLN$ \\
 random BC & $\fsr[(\mB=0)]$ & $\fsrISNmB$ &    --     \\
\end{tabular}
\endgroup
\end{table}

\subsection{Scaling limit}

In the scaling limit, with fixed aspect ratio $\rho$, the average residual free energy \eqref{eq:dFbar} at criticality becomes asymptotically equal to the Casimir amplitude $\DeltaC$ of the considered boundary conditions,
\begin{align}\label{eq:Delta^or}
\DeltaC^\OK(\rho) \simeq \delta \bar F^\OK(L,M).
\end{align}
The corresponding averaged residual of the excess free energy \eqref{eqs:Fexk} is then given by%
\bS\label{eqs:dF^ex}\begin{align}
     \delta \bar F^{\EX[,\kappa]}(L,M) &=
    \bar F^{\EX[,\kappa]}(L,M)-\bar F^{\EX[,\kappa]}_\infty(L,M), \label{eq:dFbar^ex}\\
    \bar F^{\EX[,\kappa]}_\infty(L,M) &= M(\fsr[(\kappa)] - \fsst) \label{eq:dFbar^ex_inf},
\end{align}\eS%
fulfilling
\begin{align}\label{eq:Delta^ex}
    \DeltaC^{\EX[,\kappa]}(\rho) \simeq \delta \bar F^{\EX[,\kappa]}(L,M) 
\end{align}
in the scaling limit.
Note that the bulk contributions in \eqref{eq:dFbar^ex_inf} cancel out.

Therefore, 
the disorder averaged Casimir amplitudes fulfill 
\begin{align}
\label{eq:Delta^or_2}
\DeltaC^\OK(\rho)=\DeltaC^\OST(\rho)+\DeltaC^{\EX[,\kappa]}(\rho),
\end{align}
with $\DeltaC^\OST(\rho)$ as Casimir amplitude for a system with open boundary on one side and staggered boundary condition on the other. 
Hobrecht derived a formula for the Casimir amplitude with open boundaries on both sides
\begin{align}
\label{eq:Delta^oo}
    \DeltaC^\OO(\rho)=-\frac{\pi \rho}{12}-\log\frac{(-1,\ee^{-2\pi \rho})_\infty}{(-1,\ee^{-4\pi \rho})_\infty}=\log\frac{\eta(\ii\rho)\eta(4\ii\rho)}{\eta(2\ii\rho)^2},
\end{align}
and showed that $\DeltaC^\OST(\rho)\equiv\DeltaC^\OO(\rho)$ is in the Dirichlet
surface universality class \cite{HobrechtHucht18b}. 
In \eqref{eq:Delta^oo}, $(\,\cdot\,)_\infty$ denotes the $q$-Pochhammer symbol and $\eta(\,\cdot\,)$ stands for the Dedekind eta function. 
Note that the modular identity $\eta(-1/x)=\sqrt{-\ii x}\:\eta(x)$ leads to the non-trivial symmetry for the Casimir amplitude with open boundaries
\begin{align}
\label{eq:Delta^oo_symmetry}
    \DeltaC^\OO(\rho)=\DeltaC^\OO\big(\tfrac{1}{4\rho}\big).
\end{align}
The other well known surface universality class is $(\mathrm{o},+)$, with corresponding Casimir amplitude \cite{HobrechtHucht18b}
\begin{align}
\label{eq:Delta^o+}
    \DeltaC^\OP(\rho)=\DeltaC^\OO(\rho)+\frac{\log 2}{2}-\log\frac{(\ee^{-2\pi \rho}, \ee^{-4\pi \rho})_\infty}{(-\ee^{-2\pi \rho},\ee^{-4\pi \rho})_\infty}.
\end{align}
Note that not only the boundary with all spins up, but also the one with all spins down is in this universality class due to the $Z_2$-symmetry. 
Further, boundaries like first-half spins up, second-half down belong in this class, too \cite{EvansStecki94}.

\subsubsection*{Casimir forces}

With the free energy as thermodynamic potential, a number of different quantities are now accessible. 
In the following, we look at the critical Casimir force in $\LL$ direction \cite{HobrechtHucht16a}
\begin{align}\label{eq:FC}
    \mathcal{F}_\mathrm{C}^\OK(L,M;\vec\kappa) \equiv -\frac{1}{M}\frac{\partial}{\partial L}\delta F^\OK(L,M;\vec\kappa),
\end{align}
with corresponding scaling function
\begin{align}
\label{eq:Theta_from_Fc}
    \vartheta^\OK(\rho)=-\rho \frac{\partial}{\partial \rho}\DeltaC^\OK(\rho)\simeq L M \bar{\mathcal{F}}_\mathrm{C}^\OK(L,M)
\end{align}
in the thermodynamic limit $L,M\rightarrow\infty$.
At criticality, where we restricted our calculations to, the force becomes significant for different applications, such as colloidal aggregation \cite{HobrechtHucht15a}.
 
For a description of the Casimir force, we define the Green's function $\mat G^{(\kappa)}(\mat\kappa)$ as the matrix dual \eqref{eqs:dual_pm} of $\mat\kappa \mat Q$,
\begin{align}
    \label{G_basic}
    \mat G^{(\kappa)}(\mat\kappa) \equiv (\mat\kappa \mat Q)^* = \frac
    {\mat 1 - \mat\kappa \mat Q}
    {\mat 1 + \mat\kappa \mat Q}
    = \frac 2
    {\mat 1 + \mat\kappa \mat Q} - \mat 1
    .
\end{align}
With that, the excess critical Casimir force is given by
\bS\begin{align}\label{eq:Fcex}
    \mathcal{F}_\mathrm{C}^{\EX[,\kappa]}(L,M;\vec\kappa)
    &=-\frac{1}{M}\frac{\partial}{\partial L}F^{\EX[,\kappa]}(L,M;\vec\kappa) \\
    &=-\frac{1}{4 M}\Tr\left[\bigl(\mat G^{(\kappa)}(\mat\kappa)-\mat G^{(\st)}\bigr) \, \frac{\partial \mat{Q}}{\partial L} \, \mat Q^\T\right]
\end{align}\eS%
and contains all disorder contributions.
The corresponding averaged scaling function reads
\begin{align}\label{eq:theta_ex}
    \vartheta^{\EX[,\kappa]}(\rho)=\lim_{L,M\to\infty}-\frac{L}{4}\Tr\left[\bigl(\bar{\mat G}^{(\kappa)}-\mat G^{(\st)}\bigr) \, \frac{\partial \mat{Q}}{\partial L} \, \mat Q^\T\right].
\end{align}
Note that the ensemble-averaged matrix $\bar{\mat G}^{(\kappa)} \equiv \langle\mat G^{(\kappa)}(\mat\kappa)\rangle_\kappa$ is skew-circulant and skew-sym\-metric for every well defined ensemble $\{\kappa\}$, therefore having only $M/2$ independent matrix elements.

In order to precisely calculate the excess Casimir force scaling function \eqref{eq:theta_ex}, reducing finite-size corrections becomes necessary. 
For that, the trace is taken over the eigenvalues
\begin{align}\label{eq:theta_ex_diagonalized}
    \vartheta^{\EX[,\kappa]}(\rho)=\lim_{L,M\to\infty}-\frac{L}{4}\Tr\left[\bigl(\lambda_{\bar{\mat G}^{(\kappa)}}-\lambda_{\mat G^{(\st)}}\bigr)\,\frac{\partial \lambda_{\mat{Q}}}{\partial L}\,\lambda_{\mat Q^\T}\right]
\end{align}
since all matrices can be diagonalized with the Fourier transform \eqref{eq:fourier}. 
Next, the thermodynamic limit $L,M\to\infty$ is taken for the disorder-independent part $\frac{\partial \lambda_{\mat{Q}}}{\partial L}\lambda_{\mat Q^\T}$ which leads to
\begin{align}
\label{eq:theta_ex_part_lim}
    \vartheta^{\EX[,\kappa]}(\rho) = \lim_{L,M\to\infty} \sum_{\substack{-M<\mu<M\\\mu\text{ odd}}} \frac{-2\pi\mu\rho}{\cosh^2\left(\pi\mu\rho\right)} \,
    \frac{\operatorname{Im} \bigl(\lambda_{\bar{\mat G}^{(\kappa)}}-\lambda_{\mat G^{(\st)}}\bigr)}{\zc^{-1} + \zc\tanh^2\left(\pi\mu\rho\right)} .
\end{align}
Note that the eigenvalues $\lambda_{\mat G^{(\kappa)}(\mat\kappa)}$ of every $\mat G$ are purely imaginary.

Then, the disorder averaged Casimir force scaling function is given by
\begin{align}
\label{eq:Theta_OR}
    \vartheta^{\OK}(\rho) = \vartheta^{\EX[,\kappa]}(\rho) + \vartheta^\OST(\rho).
\end{align}
For the staggered case, it reads
\begin{align}
\label{eq:Theta_oo}
    \vartheta^{\OST}(\rho)=\vartheta^{\OO}(\rho)=\frac{\pi \rho}{12} \biggl(
    \Theta_4^4(\ee^{-2\pi \rho})-\Theta_2^4(\ee^{-2\pi \rho})\biggr)
    ,
\end{align}
 as derivative of the Casimir amplitude for open boundaries \eqref{eq:Delta^oo}, with elliptic theta constants $\Theta_2(q)=2\sum_{n=0}^{\infty}(-1)^nq^{(n+1/2)^2}$ and $\Theta_4(q)=1+2\sum_{n=1}^{\infty}q^{n^2}$.

\subsection{Enumeration algorithms}

In order to simulate the with $M$ exponentially increasing many systems, two kinds of algorithms can be used. 
One algorithm based on the Sherman-Morrison-Woodbury identity\cite{ShermanMorrison1950} is used to calculate the free energy and the Casimir force for all systems, whereas the Woodbury tree algorithm defined below as well as the principal minor algorithm \cite{GriffinTsatsomeros2006} are faster, but limited to free energy calculations.
In comparison to calculating each determinant, which would have complexity $\mathcal{O}(M^3 2^M)$, both algorithms are considerably faster, with complexities $\mathcal{O}(M^2 2^M)$ for the Woodbury method, and $\mathcal{O}(2^M)$ for the Woodbury tree and principal minor algorithm, respectively.

All methods were implemented in C/C++ on CPUs \cite{MasterBueddefeld,BachelorCervellera,MasterCervellera}, and the Woodbury tree and principal minor algorithm has also been implemented on GPUs using TensorFlow \cite{BachelorOing}. 
However, due to the required high numerical precision, even with 6 NVIDIA Tesla V100S GPUs the calculations took approximately the same time as with two AMD EPYC 7742 64-Core processors.

\subsubsection*{Woodbury algorithm}

Instead of calculating the determinant $\det(\mat Q+\mat\kappa)$ repeatedly for each $\vec \epsilon$, the idea is to update the determinant consecutively based on its previous value using the Woodbury identity \cite{ShermanMorrison1950}. 
One update should correspond to a single boundary spin flip and every system should be calculated exactly once. 
These requirements are fulfilled by using the Gray code sequence \cite{WikiGrayCode} as boundary spins $\vec \epsilon$, which is a binary sequence that forms a Hamiltonian cycle while changing only one spin in each step.

By flipping one boundary spin $\epsilon_i$ at position $i$, the elements $\kappa_{\im}$ and $\kappa_i$ 
will change, leading to a slightly changed matrix $\tilde{\mat\kappa}$.
The free energy difference of one update is then given by
\bS\begin{align}
\label{eq:dF_increment_1}
    \Delta F(L,M;\vec\kappa)&= -\frac{1}{2}\log \det\frac{\mat{Q}+\tilde{\mat{\kappa}}}{\mat{Q}+\mat{\kappa}}\\
      &=-\frac{1}{2}\log \det\left(\mat{1}+\frac{1}{2}\left(\tilde{\mat{\kappa}}\mat{\kappa}-\mat{1}\right)\mat G(\mat\kappa)\right),
\end{align}\eS%
where the matrix $\frac{1}{2}\left(\tilde{\mat{\kappa}}\mat{\kappa}-\mat 1\right)$ is $-1$ at positions $(\im,\im)$ and $(i,i)$, and $0$ elsewhere.
There\-fore, the determinant can be reduced to a $2\times 2$ minor by a Laplace expansion,
\begin{align}
    \label{fe_diff}
    \Delta F(L,M;\vec\kappa)= -\log(\mat\kappa \mat G(\mat\kappa))_{\im,i},
\end{align}
where the skew symmetry of $\mat G(\mat\kappa)$ simplifies the determinant to only the $(\im,i)$ element of $\mat G(\mat\kappa)$.
\XX{
\fred{Old:}

\CC{
The Casimir force \eqref{eq:Fcex} can also be simplified with $\mat{G}$, leading to
\begin{align}
    \label{Fc_G}
     \mathcal{F}_\mathrm{C}^{\EX}(L,M;\vec\kappa)=\frac{1}{2 M}\Tr\left(\left(\frac{1}{2}\left(\mat{1}+\mat{G}\right)\mat{\kappa}-\frac{1}{\mat{Q}-\mat{1}}\right)\frac{\partial \mat{Q}}{\partial L}\right),
    \end{align}
\luca{
Alternatively:
\begin{align}
     \Delta\mathcal{F}_\mathrm{C}(L,M;\vec\kappa)=\frac{1}{4 M}\Tr\left(\left(\left(\mat{1}+\tilde{\mat{G}}\right)\tilde{\mat{\kappa}}-\left(\mat{1}+\mat{G}\right)\mat{\kappa}\right)\frac{\partial \mat{Q}}{\partial L}\right)\\
     \Delta\mathcal{F}_\mathrm{C}(L,M;\vec\kappa)=\frac{1}{4 M}\Tr\left(\left(\tilde{\mat{\kappa}}-\mat{\kappa}+\tilde{\mat{G}}\tilde{\mat{\kappa}}-\mat{G}\mat{\kappa}\right)\frac{\partial \mat{Q}}{\partial L}\right)
    \end{align}
}\fred{
\begin{align}
    \Delta\mathcal{F}_\mathrm{C} &=
     \frac{1}{4 M (\mat G)_{i,i+1}}
     \Tr\left[(\mat 1 + \mat G) \mat x
     \begin{pmatrix} 0 & 1 \\ -1 & 0 \end{pmatrix}
     \mat x^\T (\mat 1 + \mat G) \mat\kappa \frac{\partial \mat{Q}}{\partial L}\right]\\
     &=\frac{((\mat 1 + \mat G) \mat\kappa \frac{\partial \mat{Q}}{\partial L}
     (\mat 1 + \mat G))_{i,i+1}}{4 M (\mat G)_{i,i+1}}\\
     &=\frac{1}{4 M (\mat G)_{i,i+1}}
     \Tr\left[\mat g
     \begin{pmatrix} 0 & 1 \\ -1 & 0 \end{pmatrix}
     \mat g^\T \mat Q^\T \frac{\partial \mat{Q}}{\partial L}\right]\\
     &=-\frac{1}{4 M}
     \Tr\left[\Delta \mat G \mat Q^\T \frac{\partial \mat{Q}}{\partial L}\right]
\end{align}
}

The Casimir force \eqref{eq:FC} can be updated in a similar fashion, with increment
\begin{align}
    \Delta\mathcal{F}_\mathrm{C} &=
    -\frac{1}{4 M}
     \Tr\left[\Delta \mat G \, \mat Q^\T \frac{\partial \mat{Q}}{\partial L}\right],
\end{align}
where the time consuming inversion of $\mat{Q}+\mat{\kappa}$ does not have to be calculated.
Inserting $\Delta\mat G$ and performing a cyclic permutation of the trace, this expression can be further simplified to
\begin{align}
    \Delta\mathcal{F}_\mathrm{C} &=
    \frac{1}{2 M (\mat G)_{i,i+1}}
     \left(
     \mat g^\T
     \mat Q^\T \frac{\partial \mat{Q}}{\partial L}\mat g\right)_{2,1},
\end{align}
}
\fred{New:}
}

\begin{figure}[t]
    \centering
    \includegraphics[width=0.9\textwidth]{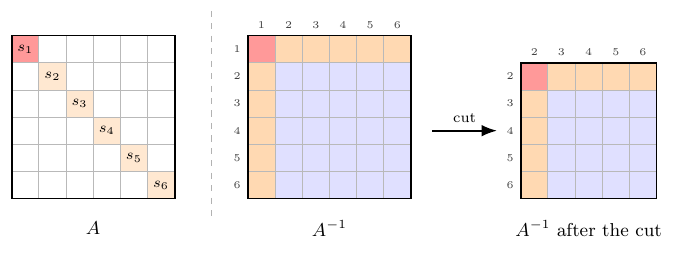}
    \caption{
    The cut applied to the tracked inverse for $M = 6$.
    The left panel shows $\mat A$, which carries the boundary spins on its diagonal; the two right panels show the tracked inverse $\mat A^{-1}$, which carries none, so that the index of the flipped spin is all the two have in common.
    As the spins are placed on the diagonal, an update of the inverse for the change of the first spin using \eqref{eq:woodbury-formula} requires an outer product from the first row and column of $\mat A^{-1}$, marked in red and orange.
    Since every subsequent flip sits at an index larger than one, neither the row nor the column is needed again, so only the blue block is carried on.
    }
    \label{fig:woodbury-cut}
\end{figure}

For an efficient update of $\mat G \mapsto \tilde{\mat G}=\mat G + \Delta\mat G$ the Woodbury identity is used. 
Due to the skew symmetry of $\mat G$, the rank-2 update of $\mat G$ can be considerably simplified. 
Defining the $M\times 1$ row vector $
|\mat g_m\rangle$ as the $m$-th row of the matrix $\mat 1 + \mat G$,
the increment $\Delta\mat G$ is simply given by
\begin{align}\label{G_update}
    \Delta \mat G
     = \frac{ |\mat g_i\rangle\langle \mat g_\im| - |\mat g_\im\rangle\langle \mat g_i|}{(\mat G)_{\im,i}}.
\end{align}
The Casimir force \eqref{eq:FC} is updated in a similar fashion according to $\mathcal{F}_\mathrm{C} \mapsto \tilde{\mathcal{F}}_\mathrm{C}=\mathcal{F}_\mathrm{C} + \Delta\mathcal{F}_\mathrm{C}$, with increment
\begin{align}
    \Delta\mathcal{F}_\mathrm{C} &=
    -\frac{1}{4 M}
     \Tr\left[\Delta \mat G \, \mat Q^\T \frac{\partial \mat{Q}}{\partial L}\right].
\end{align}

\subsubsection*{Woodbury tree algorithm}

The scheme described above visits the $2^M$ boundary spin configurations $\vec\epsilon$ one after another along the Gray code, updating $\mat G$ at every step.
If, as here, all configurations are required anyway and the updated $\mat G$ is not needed beyond the determinants it supplies, this traversal can be reorganised into a tree.
The resulting Woodbury tree algorithm is an alternative scheme to the Woodbury technique.
It can be very useful where spin-dependent determinants have to be calculated and the complete inverse of the matrix is not required. It bears major similarities to the principal minor technique \cite{GriffinTsatsomeros2006}, but was developed independently.

\begin{figure}[t]
    \centering
    \includegraphics[width=0.75\textwidth]{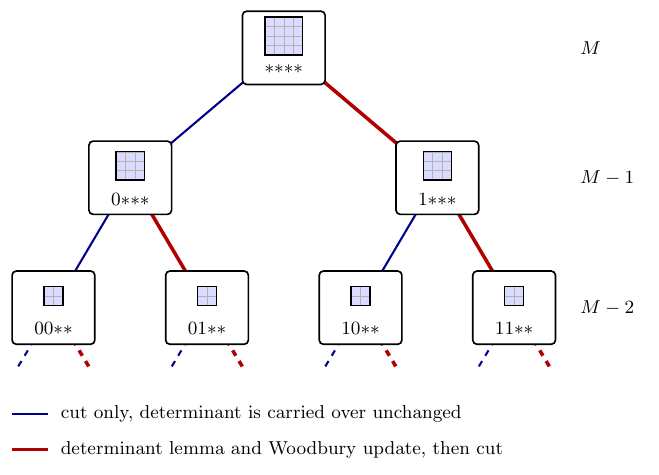}
    \caption{
    Scheme of the Woodbury tree algorithm for four boundary spins.
    Every level of the tree decides one spin, so that a digit denotes an already fixed spin and $\ast$ a spin that is still free.
    At the root we start from the inverse of the base matrix, obtained once explicitly.
    Descending to the left leaves the spin at $0$, so that only the row and the column of that spin are cut off.
    Descending to the right flips the spin to $1$, for which the matrix determinant lemma \eqref{eq:matrix-determinant-lemma} and the Woodbury identity \eqref{eq:woodbury-formula} are applied before the same cut.
    The leaves carry the determinants of all $2^M$ spin configurations in binary order.
    }
    \label{fig:woodbury-tree}
\end{figure}

The idea comes from the necessity of the application of the Woodbury identity
\begin{equation}
    (\mat A + \mat u \mat C \mat v)^{-1} = \mat A^{-1} - \mat A^{-1} \mat u ( \mat C^{-1} + \mat v \mat A^{-1} \mat u)^{-1} \mat v \mat A^{-1} \, \text{,}
    \label{eq:woodbury-formula}
\end{equation}
and from the matrix determinant lemma
\begin{equation}
    \det(\mat A + \mat u \mat C \mat v) = \det(\mat A) \det(\mat C^{-1} + \mat v \mat A^{-1} \mat u) \det(\mat C) \,\text{,}
    \label{eq:matrix-determinant-lemma}
\end{equation}
which describe the change of the matrix inverse $\mat A^{-1}$ and the determinant $\det(\mat A)$ for a local change in the original matrix $\mat A$ described by $\mat A + \mat u \mat C \mat v$, with the projectors $\mat u = \mat v^\T$.
In the present case $\mat A = \mat Q + \mat\kappa$, and $\mat A^{-1}$ is the tracked matrix denoted $\mat G$ above.
Since flipping the boundary spin $\epsilon_i$ alters the two diagonal entries $\kappa_\im$ and $\kappa_i$, the change matrix $\mat C$ is of size $2\times 2$ and $\mat u$ is of size $M\times 2$, so that each step is a rank-2 update.

The main advantage of the Woodbury tree algorithm compared to the brute force application of the above formulas is speed.
The core idea is that the rows and columns of $\mat A^{-1}$ whose index lies at or below the last spin change never have to be updated again, as they do not enter the calculation of any of the following determinants.
As such, the calculation-expensive outer products in \eqref{eq:woodbury-formula} can be reduced by cutting off all previous spins, as shown in Fig.~\ref{fig:woodbury-cut}.

This allows us to apply the Woodbury technique in a tree like fashion, as shown in Fig.~\ref{fig:woodbury-tree}.
As an additional advantage, the usage of the Woodbury tree algorithm places all determinants in a binary ordering at the bottom of the tree, so that the Gray code is no longer required to order the configurations.

Passing through all $2^M$ spin configurations sequentially along the Gray code requires one full-matrix outer product per configuration and therefore scales as $\mathcal{O}(2^M M^2)$.
In the tree, an update is only required when descending to the right, and at level $i$ the tracked inverse has already shrunk to a size of $M-i$, which leaves $\sum_{i=1}^{M} 2^{i-1} (M-i)^2 = 3 \cdot 2^M - M^2 - 2M - 3$, so that the tree scales as $\mathcal{O}(2^M)$ and performs a constant amount of work per spin configuration.
The speed-up over the sequential scheme therefore grows as $M^2/3$, which amounts to a factor of $300$ for $M = 30$.
The inversion at the top of the tree happens once and is therefore negligible.

\XX{Finally, we compare the recursive pseudo-code of the Woodbury tree algorithm with the principle minors algorithm \cite{GriffinTsatsomeros2006}.
}

\CC{\subsubsection*{Principal minor algorithm}
\luca{old:}
The principal minor algorithm is based on the publication 
by Griffin and Tsatsomeros \cite{GriffinTsatsomeros2006} and calculates the $2^M$ principal minors of a given $M\times M$ matrix in $\mathcal{O}(2^M)$ operations. 
Starting with $\mat G(\mat 1)$, the algorithm calculates its principal minors which correspond directly to the update increments for the free energy \eqref{eq:dF_increment_1}. 
To avoid the Pivot-corrections mentioned in \cite{GriffinTsatsomeros2006}, we can use in our case the Schur-complement of $\mat G(\mat 1)$ to get the same free energy increments in less computation time. 

\fred{Zu kurz}
\luca{new}

As the principal minors of our propagator $\mat G(\mat\kappa)$ are connected directly to the update increments for the free energy \eqref{eq:dF_increment_1}, an even more efficient algorithm, than Woodbury can be used in this case.
The principal minor algorithm is based on the work 
of Griffin and Tsatsomeros \cite{GriffinTsatsomeros2006}, where they developed an algorithm that calculates all $2^M$ principal minors of a given $M\times M$ matrix within a computation time of $\mathcal{O}( 2^M)$.

There, the algorithm follows a binary tree, either by removing the first row and column of the matrix or by taking its Schur-complement with respect to the first matrix element. Then, the first element of the new matrix is a principal minor of the original matrix and the algorithm repeats.

Here, the matrix reduction relates to a single boundary spin flip at site $k$ and the Schur complement to flipping the first $k$ boundary spins, where $k$ is the depth in the binary tree. This scheme is depicted in figure \ref{fig:Binary-Tree}, where paths to the left (orange) perform a matrix reduction and paths to the right a Schur complement calculation.
\begin{figure} 
    \includegraphics{figures/Tree-Abb-final.pdf}\hfill
\label{fig:Binary-Tree}
\caption{Algorithm representation as binary tree. Each node shows the actual boundary spin state and paths to the left or to the right represent matrix reductions or Schur complement calculations, respectively. The starting state is here chosen to be $\mat\kappa=\mat 1$.}
\end{figure}
Note that for $k=1$, both paths yield the same spin state and therefore only one side of the tree has to be calculated. We choose the right path in order to get diagonal elements that are non-zero. With that, Pivot-corrections which are mentioned in the original work \cite{GriffinTsatsomeros2006} can be avoided. Furthermore, the $Z_2$-symmetry of the Ising model is used, which halves the number of spin states that need to be calculated.
In principle the start configuration $\mat\kappa$ can be any boundary spin state, but it seems to be reasonable to start with $\mat\kappa =\pm\mat 1$ as the propagator $\mat G (\pm\mat1)$ of these states can be calculated analytically.
Also, the algorithm can be adjusted to calculate $\mat G(\kappa)$ and $\mat G(-\kappa)$ simultaneously in a way that fewer time intensive Schur-complements have to be calculated \luca{wie war das nochmal genau? Nur du hattest das ja implementiert, ich weiss das nichtmehr genau}.

}
\section{Results}

\subsection{Scaling form of the free energy histograms}\label{sec:histo}

\newcommand{\MMM}{\mathcal{M}}

We start with a motivation for the log-normal form and first focus on the infinite long cylinder with $\rho,L\to\infty$:
Define the marginal distribution of the residual excess free energy DOTS $\delta\omega^{\EX}$ from \eqref{eq:dots^ex},
\begin{align}\label{eq:deltaomega^ex}
    \delta\omega^{\EX}(L,M;\delta F^{\EX}) &= \int\! \dd \MB \, \omega^{\EX}(L,M;F^{\EX},\MB) \, ,
\end{align}
with $F^{\EX} = (f^{\OK}-f^{\OST})M+\delta F^{\EX}$ from \eqref{eqs:Fexk}, such that
\begin{align}
    \delta \bar F^{\EX}(L,M) &= \int\! \dd \phi \, \phi \, \delta\omega^{\EX}(L,M;\phi) \, .
\end{align}
We know that 
\begin{align}\label{eq:DeltaC_of_deltaF}
\DeltaC^{\EX}(\rho) \simeq
     \langle \delta F^{\EX}(L,M;\vec\kappa) \rangle_\kappa 
\end{align}
is universal. 
We now argue that the scaling form of the DOTS $\delta\omega^{\EX}(L,M;\delta F^{\EX})$ is universal, too.

Numerical evidence shows that $-F^{\EX}/M$ (and therefore also $-\delta F^{\EX}/M$) is asymptotically log-normal distributed, with cutoff $f_0$, and that the width scales like $\sigma(M)\simeq \sigma_0/(f_1 M)^{1/2}$, such that
\begin{align}\label{eq:deltaOmega_1}
    \delta\omega^{\EX}(L,M;\delta F^{\EX}) \simeq \frac{1}{f_2 M}\mathcal{LN}\left(\mu(L,M),\sigma(M);\frac{f_0-\delta F^{\EX}/M}{f_2}\right),
\end{align}
with yet unknown constants $f_{1,2}$.
The first moment of $\mathcal{LN}(\mu,\sigma;x)$ is known to be $\langle x\rangle=\ee^{\mu+\sigma^2/2}$. 
Expanding around $M=\infty$ and comparing to the scaling prediction \eqref{eq:DeltaC_of_deltaF} leads to the conclusion that $\mu(L,M)\simeq \mu_0(\rho)/(f_1 M)$, with constants $f_2=f_1=f_0$, such that the resulting scaling form reads
\begin{align}\label{eq:deltaomega^ex_2}
    \delta\omega^{\EX}(L,M;\delta F^{\EX}) \simeq \frac{1}{\MMM} \, \mathcal{LN}\left(\frac{\mu_0(\rho)}{\MMM},\frac{\sigma_0}{\sqrt{\MMM}};1-\frac{\delta F^{\EX}}{\MMM}\right),
\end{align}
with dimensionless 
scaling variable $\MMM=
f_0 M$, such that  the excess Casimir amplitude becomes
\begin{align}
    \DeltaC^{\EX}(\rho) =
 -\left(\mu_0(\rho)+\frac{1}{2}\sigma_0^2\right),
\end{align}
i.e., the Casimir amplitude has two contributions, one homogeneous from the average $\mu$, and one from the disorder fluctuations.
The remaining unknown $f_0$ can be adjusted to compensate for remaining leading order corrections. 
For $\rho=\infty$ we find $f_0 \approx 0.07$.

\XX{
\newcommand{\ewk}[1]{\big\langle #1 \big\rangle_\kappa}
\newcommand{\dF}{\delta F}

\subsubsection*{New justification of log-normal free energy distribution (copied from end)}

In this section we will motivate a mechanism...
For simplicity we let $L\to\infty$ and drop the argument $L$ as well as the superscript $\OR$.
We are interested in the distribution of residual free energies around the Casimir amplitude. 
The average residual free energy $\delta\bar F(M)$ from \eqref{eq:dFbar} fulfills
\begin{align}
    \delta\bar F(M) 
    &
     = \ewk{\delta F(M;\vec\kappa)} \\
    &= -\ewk{\log \delta Z(M;\vec\kappa)} \\
    &= -\frac{\partial}{\partial n} \ewk{\delta Z(M;\vec\kappa)^n} \Big|_{n\to 0} \\
    &= \lim_{n\to 0} \tfrac 1 n \ewk{1 - Z(M;\vec\kappa)^n}
\end{align}
Let $\Theta=\ee^{-\Delta}$
\begin{align}
Z(M;\vec\kappa) &= \Theta Z_s(M;\vec\kappa) \\
    \fs &= \lim_{M\to\infty} \tfrac 1 M (\bar F(M)-\Delta) \\
    &= \lim_{M\to\infty}\lim_{n\to 0} \tfrac 1 {M n}\ewk{1 - Z_s(M;\vec\kappa)^{n/M}}
\end{align}

\subsubsection*{Alternative argumentation for marginal distribution}

\newcommand{\ham}{\mathrm{ham}}

Drop $\OR$ and let $L\to\infty$, fix notation later.

\noindent
We are interested in the distribution of residual free energies around the Casimir amplitude. 
We start with the residual free energy $\dF$ from \eqref{eq:DeltaC_of_deltaF} and define the quench fluctuations
\begin{align}
    \Phi(M;\vec\kappa) &= \delta F(M;\vec\kappa)-\DeltaC
    \intertext{with average}
    \bar\Phi(M)=\ewk{\Phi(M;\vec\kappa)} &= \delta\bar F(M) - \DeltaC + o(1) \\
    \intertext{fulfilling}
    \lim_{M\to\infty}\ewk{\Phi(M;\vec\kappa)} &= \lim_{M\to\infty} \bar\Phi(M) = 0
\end{align}
To leading order,
\begin{align}
    \ewk{\Phi(M;\vec\kappa)^2} = \hat\sigma^2 M + O(1)
\end{align}
with nonuniversal variance $\hat\sigma^2$ ($\hat\sigma^2_{\iso} \approx 0.01$, $\hat\sigma^2_{\hl} \approx 0.0625$).

The quench fluctuations density $\phi=\Phi/M$ gives
\begin{align}
    \ewk{\phi(M;\vec\kappa)^2} = \frac{\hat\sigma^2}{M} + O(M^{-2})
\end{align}

$\phi$ has distribution
$$
\bar\phi(M) = \int \phi \, \omega(M;\phi) \, \dd \phi,
$$
such that the width of $\omega(M;\phi) \sim 1/\sqrt{M}$.
\newcommand{\ZZ}{\Psi}
The quenched average of $\Phi$ \XX{\luca{der Variablenname $\Phi$ wird bereits benutzt}} is (define corresponding partition function $\ZZ=\ee^{-\Phi}$)%
\bS\begin{align}
    \bar\Phi(M)
    &= \ewk{\Phi(M;\vec\kappa)} \\
    &= -\ewk{\log \ZZ(M;\vec\kappa)} \\
    &= -\frac{\partial}{\partial n} \ewk{\ZZ(M;\vec\kappa)^n} \Big|_{n\to 0} \\
    &= \lim_{n\to 0} \tfrac 1 n \big[1 - \ewk{\ZZ(M;\vec\kappa)^n} \big] \\
    &= \lim_{n\to 0} \tfrac 1 n \big[1 - \ewk{\ee^{-n \, \Phi(M;\vec\kappa)}} \big]
\end{align}\eS%
\newcommand{\zz}{\psi}
$\zz=\ee^{-\phi}$, $Z \propto \zz^M = \ee^{-M \phi}$
\bS\begin{align}
    \bar\phi(M)
    &= \ewk{\phi(M;\vec\kappa)} \\
    &= -\ewk{\log \zz(M;\vec\kappa)} \\
    &= -\frac{\partial}{\partial n} \ewk{\zz(M;\vec\kappa)^n} \Big|_{n\to 0} \\
    &= \lim_{n\to 0} \tfrac 1 n \big[1 - \ewk{\zz(M;\vec\kappa)^n} \big] \\
    &= \lim_{n\to 0} \tfrac \nu n \ewk{1-\ee^{-\frac{n \, \phi(M;\vec\kappa)}{\nu}}} \\
    &= \lim_{n\to 0} \tfrac \nu n \ewk{1-\ee^{-\frac{n \, \Phi(M;\vec\kappa)}{\nu M}}} 
\end{align}\eS%
with constant $\nu$.
} 
\XX{
\begin{figure}
    \centering
    \includegraphics[width=0.48 \textwidth]{figures-dist/dist-1.pdf}\hfill
    \includegraphics[width=0.48 \textwidth]{figures-dist/dist-2.pdf}\\
    \includegraphics[width=0.48 \textwidth]{figures-dist/dist-3.pdf}\hfill
    \includegraphics[width=0.48 \textwidth]{figures-dist/dist-4.pdf}\\
    \includegraphics[width=0.48 \textwidth]{figures-dist/dist-5.pdf}\hfill
    \includegraphics[width=0.48 \textwidth]{figures-dist/dist-6.pdf}\\
    \includegraphics[width=0.48 \textwidth]{figures-dist/dist-7.pdf}\hfill
    \includegraphics[width=0.48 \textwidth]{figures-dist/dist-8.pdf}
    \caption{Some figures}
    \label{fig:enter-label}
\end{figure}

} 

\subsection{Critical random surface free energy}\label{sec:fsr}

\begin{table}[b!]
\centering
\caption{Critical surface free energy densities $\fsr[(\,\cdot\,)](\zc)$ of the quenched random boundary for different disorder ensembles and critical couplings $\zc$. 
}
\begin{tabular}{c|D{.}{.}{14}|D{.}{.}{14}|D{.}{.}{14}} 
Method & 
\multicolumn{1}{c|}{$\fsr(\zc^\iso)$} &
\multicolumn{1}{c|}{$\fsr[(\mB=0)](\zc^\iso)$} &
\multicolumn{1}{c}{$\fsr(\zc\to1)$} \\
\hline\hline
classical fit       & 0.00266(11\pm30)  & 0.002(71\pm12)   & 0.26536(91\pm32)  \\
$\Psi$-method + fit & 0.002662(6\pm7)   & 0.00266(48\pm24) & 0.26537(08\pm22)  \\
Aitken's            & 0.002662(10\pm18) & 0.0026(5\pm6)    & 0.265369(2\pm7)   \\
Levin-U             & 0.002662(0\pm4)   & 0.00266(37\pm32) & 0.265369(15\pm33) \\
Bulirsch-Stoer      & 0.0026628(2\pm5)  & 0.00266(3\pm1)   & 0.2653711(3\pm5) \\
\hline
average             & \fsrISN  & \fsrISNmB & \fsrHLN \\ 
maximum $M$&\multicolumn{1}{c|}{$48$}&\multicolumn{1}{c|}{$40$}&\multicolumn{1}{c}{$54$}\\ 
\end{tabular}
\label{tab:fsr}
\end{table}

For the determination of the leading surface divergence $\fsr M$ from \eqref{eq:fsr}, we calculate the finite differences of \eqref{eqs:Fexk} in $M$, leading to a sequence converging towards 
\begin{align}
\label{eq:Finite_diff_fe}
\fsr[\EX] =\fsr-\fsst=\lim_{M\to\infty}\frac{\bar F^{\EX}(L\to\infty,M+1)-\bar F^{\EX}(L\to\infty,M-1)}{2}
\end{align}
in the thermodynamic limit. 
Note that here $L\to\infty$ is used as these semi-infinite systems show the smallest lattice corrections. 

Beside a classical polynomial fit including logarithmic corrections, we apply different extrapolation methods for numerically exact data.
When using a sequence $a_n$ of exactly calculated values, the convergence can be accelerated with different extrapolation methods. In this work, we use (i) Aitken's delta squared process \cite{Aitken1927}
\begin{align}
\label{eq:Aitken}
S[a_n]=\frac{a_{n+1}a_{n-1}-a_n^2}{a_{n+1}-2a_n+a_{n-1}}\,,
\end{align}
(ii) the Levin-U transform \cite{Levin1972}
\begin{align}
\label{eq:LevinU}
    u_{k,b}^{(n)} [a_n]=\frac{\sum_{j=0}^k(-1)^j\binom{k}{j}\frac{(b+n+j)^{k-2}}{(b+n+k)^{k-1}}\frac{a_{n+j}}{a_{n+j}-a_{n-1+j}}}{\sum_{j=0}^k(-1)^j\binom{k}{j}\frac{(b+n+j)^{k-2}}{(b+n+k)^{k-1}}\frac{1}{a_{n+j}-a_{n-1+j}}},
\end{align}
where we use in the following $b=1$, and $k$ up to $k=12$, as well as (iii) the generalized Bulirsch-Stoer method \cite{HenkelSchutz1988,BulirschStoer1964}, with $\omega=0.8$.

Another method to get closer to the limit $M\to\infty$ is by canceling out higher order corrections. 
By assuming that these corrections follow some type of Taylor expansion like $\sum_k c_k M^{-k}$ in the limit $M\to\infty$, the $k$-th term can be canceled out by applying the generalized difference operator
\begin{align}
\label{eq:Psi_def}
    \Psi_k[a_n]=\frac{(n-1)^k a_{n+1}-(n+1)^ka_{n-1}}{(n-1)^k-(n+1)^k}
\end{align}
to the sequence $a_n$.
Note that in terms of the free energy, $a_n=F(M)$ and $n=M$. 
Also, equation \eqref{eq:Finite_diff_fe} can be expressed as $\fs^{\EX}=\Psi_0[\bar F^{\EX}(L\to\infty,M)]$.

In order to determine the averaged surface free energy densities $\fsr[(\,\cdot\,)]$ shown in Table \ref{tab:fsr}, the computed free energies are either extrapolated with a classical fit method or with one of the four methods 
mentioned above.
We find $\fsr[(\mB=0)]=\fsr$ independent of the disorder ensemble, where the deviation between different methods is slightly larger in the ensemble $(\mB=0)$ than in the ensemble (r). 
\XX{Note that $\fsr[(\mB=0)]=\fsr$ is not due to universality, as then $\fs[(o)]$ and $\fsst$ would also be equal to $\fsr$. 
To understand this, we want to refer to $\omega^{\EX}$ in Fig.~\ref{fig:DOTS^ex} again. 
There, the maximum of $\omega^{\EX}$ is $\fsr-\fsst$ in the thermodynamic limit $L,M\to\infty$. Due to the $\mathbb{Z}_2$-symmetry it is located at $\mB=0$.
In contrast to the residual free energy distribution discussed in Sec.~\ref{sec:histo}, the leading surface free energy terms show Gaussian fluctuations, ...
}

\subsection{The Casimir amplitude}

The computationally obtained free energies can be transformed with the operator $\Psi$ to a sequence that converges towards the Casimir amplitude
\begin{align}
\label{eq:Delta_through_psi}
    \lim_{L,M\to\infty}\Psi_1[\bar F^{\EX}(L,M)]=\DeltaC^{\EX}(\rho)
\end{align}
in the thermodynamic limit with the advantage that the surface divergence cancels out exactly.

Further, the four methods classical fit, Aitken transform \eqref{eq:Aitken}, Levin-U transform \eqref{eq:LevinU} and a combination of the $\Psi$-method \eqref{eq:Psi_def} and fit are used to calculate the Casimir amplitudes for different ensembles.

\begin{figure}
	\centering
	\includegraphics[width=1\textwidth]{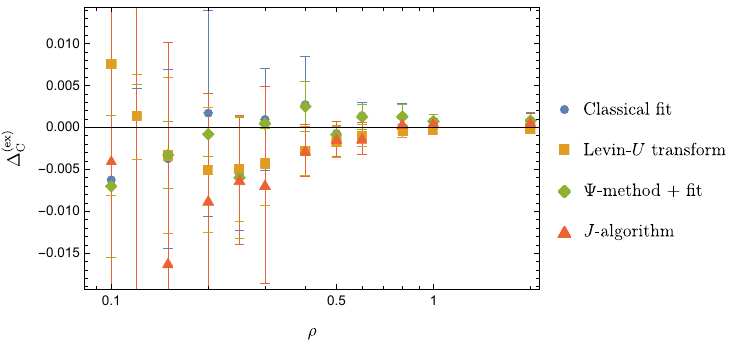}
	\caption{Excess Casimir amplitude \eqref{eq:Delta^ex} for $\zc=\zc^\iso$ and different values of $\rho$. The value at $\rho\to\infty$ is indistinguishable from zero in this representation and is therefore not shown.
    }
	\label{fig:Delta^ex}
\end{figure}

In figure \ref{fig:Delta^ex}, the excess Casimir amplitude \eqref{eq:Delta^ex} is plotted in dependence of aspect ratio $\rho$.
The values from different methods scatter around zero, where the deviation shows a minor logarithmic dependence on $\rho$. 
Most likely, this is an error occurring from finite-size effects for small $\rho=L/M$ and $L$ as $M$ is fixed due to the computation time.
Further, Diehl and Nüsser \cite{DiehlNuesser1990} derived a Harris criterion for surfaces with conclusion that surface disorder with mean zero becomes irrelevant in the Ising universality class for arbitrary dimensions $d\neq2$. 
For the case $d=2$, there could not be made a prediction, but our data \ref{fig:Delta^ex} lead to the hypothesis that $\DeltaC^{\EX}(\rho)\equiv\DeltaC^\OR(\rho)-\DeltaC^\OO(\rho)=0$.

For $\rho\to\infty$, we find for isotropic couplings
\begin{align}
     \zc=\zc^\iso:\;\lim_{\rho\to\infty}\DeltaC^{\EX}(\rho)=(2.8\pm5)\times 10^{-4}
\end{align}
and in the Hamiltonian limit 
\begin{align}
    \zc\to1:\;\lim_{\rho\to\infty}\DeltaC^{\EX}(\rho)=(1.9\pm3)\times 10^{-5}
\end{align}
via combination of $\Psi$-method and fit. 
Note that the Casimir amplitude is a universal quantity and therefore cannot depend on the coupling. 
Rather, both Casimir amplitudes are compatible with zero, and the deviations stem from finite-size effects. 
This systematic error is expected to be minimal in the limit $\rho\to\infty$ and increases monotonically with decreasing $\rho$.

\subsection{Casimir force scaling function}
\begin{figure} 
    \centering
    \includegraphics[height=0.5\textwidth]{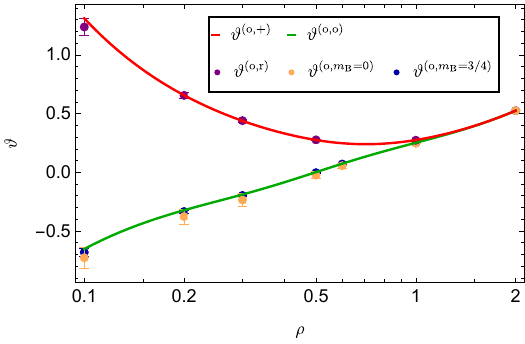}\hfill
\caption{The dots represent Casimir force scaling functions that are calculated with the $\Psi$-method. 
Lines are analytically known scaling functions, see text.
\label{figs:forces}
}
\end{figure}

Finally, we discuss the Casimir force scaling function $\vartheta$ in terms of aspect ratio $\rho$ and given ensemble.
To calculate the force, we need to use the Woodbury algorithm. As it is slower than the Woodbury tree algorithm, our maximum system size is reduced to $M=40$. Nevertheless, we achieve about the same precision for $\vartheta$ due to an analytic reduction of finite-size effects, see \eqref{eq:theta_ex_part_lim}. 
 Figure \ref{figs:forces} shows $\vartheta^\OR,\:\vartheta^\OmB$ and $\vartheta^\OmBdv$, with analytic known $\vartheta^\OO$ \eqref{eq:Theta_oo} and $\vartheta^\OP $\eqref{eq:Delta^o+},\eqref{eq:Theta_from_Fc} as reference. Here, we use a combination of $\Psi$-method and fit to get $\vartheta$.
 We find minor deviations for $\vartheta^\OR(\rho)$ and $\vartheta^\OmB$ from $\vartheta^\OO(\rho)$. We suspect some slow decaying correction terms, potentially logarithmic ones. Nonetheless, the random ensemble $\OR$ and the one with fixed magnetisation to zero $\OmB$ seem to be in the $\OO$ universality class.
 Further, we calculated $\vartheta^\OmBdv$.
 As predicted for symmetry breaking BCs from field theory 
 it seems to be in the $\OP$ universality class. 
Again, there is a larger deviation for the smallest value of $\rho=0.1$ which is due to finite-size corrections.

\CC{Next, the Casimir force scaling function will be discussed for arbitrary $\rho$ and different ensembles. 
Although the algorithm that is used to calculate the force, is slower than for the free energy and thus reducing the maximum system size to $M=40$, we achieved roughly the same precision for the Casimir force scaling function due to an analytic reduction of finite-size effects, see \eqref{eq:theta_ex_part_lim}. 
The determination of Casimir force scaling functions $\vartheta$ is done again via combination of $\Psi$-method and fit.
Figure \ref{fig:force_or_mB0} shows $\vartheta^\OR$ and $\vartheta^\OmB$, with analytic known $\vartheta^\OO$ \eqref{eq:Theta_oo} as reference.

The deviation of $\vartheta^\OR(\rho)$ from $\vartheta^\OO(\rho)$ is again visible for small $\rho\lesssim 0.3$ and is attributed to increasingly strong finite-size effects. 
The force scaling function $\vartheta^\OmB(\rho)$ deviates more from $\vartheta^\OO(\rho)$ than $\vartheta^\OR(\rho)$. 
On the one hand, the ensemble $(\mB=0)$ already showed larger finite-size corrections for the surface free energy density, see Tab.~\ref{tab:fsr}, which makes it likely that this deviation is in fact an error that cannot be estimated. 
On the other hand, this could be a new surface universality class as the ensemble (r) can be imagined as a grand canonical ensemble, whereas the ensemble $(\mB=0)$ would represent a canonical ensemble.

Additionally, we find the sign change of the Casimir force scaling function at $\rho_0^\OmB\approx0.52$, in comparison to the open cylinder $\rho_0^\OO=\rho_0^\OR=1/2$.
The Casimir force scaling function for the ensemble $(\mB=3/4)$ is shown in figure \ref{figs:forces}b.

As predicted from field theory, the scaling function $\vartheta^\OmBdv$ seems to be in the $\OP$ universality class in realm of the error margin. 
Again, there is a larger deviation for the smallest value of $\rho=0.1$ which is due to finite-size corrections.
}

\section{Conclusions}

We presented an exact solution for the free energy of a square lattice Ising cylinder with one arbitrary quenched boundary. There, we focused our work on surface disorder effects for different couplings at criticality.

The probability density function of free energies in the disorder ensemble (r) of all $2^M$ possible boundaries turns out to behave like a log-normal distribution, a property that could also be shown recently in the framework of a random-matrix-theory approach of a closely related model \cite{GluGuHu2024}.

Further, multiple quantities are calculated in the thermodynamic limit by using various sequence acceleration techniques to reduce finite-size corrections.

The surface free energy densities are independent on the ensemble being (r) or $(\mB=0)$. We showed this for isotropic ($\zc^\iso$) and maximal anisotropic ($\zc\to1$) couplings with a precision of 6 digits after the decimal point.
\XX{
\luca{This seems rather untypical as the universality argument does not hold for the surface free energy density.} 
\fred{Ich finde das nicht untypical. Koennte mit self-averaging zu tun haben.}
By comparing the values for the (r) ensemble with well known boundaries like staggered $(\st)$ or open (o), we find a clear distinction for arbitrary anisotropies. Only in the limit of $\zc\to0$, the surface decouples. There, every surface free energy density becomes indistinguishable regardless of its boundary condition.}

Unlike the free energy surface density, the Casimir amplitude, the finite-size scaling function of the free energy, is the same for open boundaries as it is for the disorder ensemble (r). 
Therefore, we verify the predictions from conformal field theory with our exact solution of a quenched boundary for $\rho\to\infty$ and see slight deviations in dependence of $\rho$, possibly due to finite-size corrections.

With our calculations of the Casimir force scaling functions for the ensembles (r) and $(m_\mathrm{B}=0)$ we support this statement even further. 
Finally, we proved for the Casimir force scaling functions that the ensemble $(\mB=3/4)$ is equal to the well known $(+)$ boundary. This highlights the hypothesis of only two surface universality classes, being (o) and (+), if the other boundary condition is open (o). Then, every surface with fixed $\mB\neq0$ is in the (+) universality class.

\section*{Acknowledgements}


\paragraph{Funding information}
This work was partially supported by the Deutsche Forschungsgemeinschaft (DFG, German Research Foundation) through project 523060084 (TraSPI) and project 278162697 (SFB 1242).


\addtocontents{toc}{\protect\setcounter{tocdepth}{1}}
\begin{appendix}
\numberwithin{equation}{section}
\section{Appendix}

\CC{

\fred{REMOVED STUFF}

The analytical calculation of $\mat Q$ from \eqref{eq:Q_0} leads to the eigenvalues
\begin{align}
\label{eq:Q_EVs}
    \lambda_{\mat Q,\mu}=\left( \FF^\dag \mat Q \FF \right)_{\mu,\mu} =-\ee^{\ii\varphi}\frac{1+(\sin\omega+\ii z \cos\omega)\tanh{\left(L \gamma\right)}}{1+(\sin\omega-\ii z \cos\omega)\tanh{\left(L \gamma\right)}}
\end{align}
for even $M$ and arbitrary positive 
\fred{$L=\rho r_\xi M$}, see \eqref{eq:rxi_def}. 
The three angles introduced in \eqref{eq:Q_EVs} are all dependent on $\mu$ and given by
\begin{align}
\label{eq:angles}
    \gamma=\sgn(\varphi)\arccosh\left(t_+z_+-t_-z_-\cos\varphi\right)\\
    \omega=\arctan\left( \frac{t_+z_--t_-z_+\cos\varphi}{t_-\sin\varphi} \right)\\
    \varphi=\frac{\pi \mu}{M}.
\end{align}
[...] The definition of $\omega$ serves aesthetic purpose and fulfills the relation $\ii\tan\omega=\cosh\theta$, with $\theta$ as relevant angle for the Ising rectangle \cite{Hucht21a}.

For the Hamiltonian limit $\lim_{t,z\nearrow1}$, $\mat Q$ has to be series expanded around $\delta\to 0$, with $t,z=1-\delta$, leading to
\begin{align}
\label{eq:Q_EVsHL}
\lim_{\delta\to 0} \lambda_{\mat Q,\mu} = \frac{ \ee^{\ii\varphi}+\ii \ee^{\ii\frac{\varphi}{2}} \tanh\left(2 \rho M \sin\frac{\varphi}{2}\right)  }{ -1+\ii \ee^{\ii\frac{\varphi}{2}} \tanh\left(2 \rho M \sin\frac{\varphi}{2}\right)}+\mathcal{O}\left(\delta\right)^2.\XX{\\
\luca{=1+\frac{1+ \ee^{\ii\varphi} }{ -1+\ii \ee^{\ii\frac{\varphi}{2}} \tanh\left(2 \rho M \sin\frac{\varphi}{2}\right)}+\mathcal{O}\left(\delta\right)^2}.\notag
}
\end{align}
In order to calculate the critical Casimir force in \eqref{eq:Fcex}, the derivative $\frac{\partial \mat Q}{\partial L}$ is needed, which can be expressed as
\label{eq:dQdL_EVs}
\begin{align}
   \frac{\partial \lambda_{\mat Q,\mu}}{\partial L} =\left( \FF^\dag \frac{\partial \mat Q}{\partial L} \FF \right)_{\mu,\mu} =\frac{2\ii\ee^{\ii\varphi} \gamma z \cos\omega }{\cosh^2\left(L\gamma\right)}\left(\frac{1}{\ii+(\ii\sin\omega+ z \cos\omega)\tanh\left(L \gamma\right)}\right)^2
\end{align}
for arbitrary $t,z$ and as
\begin{align}
\label{eq:dQdL_EVsHL}
    \lim_{\delta\to 0} L\frac{\partial \lambda_{\mat Q,\mu}}{\partial L} &= \frac{2\ii \rho M\ee^{\ii\varphi}  \sin\varphi  }{\cosh^2\left(2 \rho M \sin\frac{\varphi}{2}\right)}\left(\frac{1}{\ii+\ee^{\ii\frac{\varphi}{2}}\tanh\left( 2 \rho M \sin\frac{\varphi}{2} \right)}\right)^2+\mathcal{O}\left(\delta\right)^2
\end{align}
for the Hamiltonian limit. 
Note that the eigenvalue $\frac{\partial \lambda_{\mat Q,\mu}}{\partial L}$ has to be multiplied with $L$ first before taking the limit, in order to get a finite result.

}

\subsection{Surface free energies}\label{sec:fs}
The factorization of the partition function \eqref{eq:Zor} leads to the free energy decomposition%
\bS\begin{align}
    F^\OR(L,M;\vec\kappa) &= F^{\mathrm{(o,n)}}(L,M) + F^{\mathrm{(n,r)}}(M;\vec\kappa), \\
    \intertext{with}
    \label{eq:Fon}
    F^{\mathrm{(o,n)}}(L,M) &= -\frac12 \log(C\det\mat P), \\
    \label{eq:Fnr}
    F^{\mathrm{(n,r)}}(M;\vec\kappa) &=-\frac12 \log\det(\mat Q + \mat\kappa),
\end{align}\eS%
where we introduced an artificial surface $\mathrm{(n)}$ between the matrices $\mat P$ and $\mat Q+\mat\kappa$.
The leading contributions of the two terms are therefore%
\bS\begin{align}
    \label{eq:F∞on}
    F_\infty^{\mathrm{(o,n)}}(L,M) &= LM\fb + M(\fs[(o)] + \fs[(n)]) \\
    \label{eq:F∞nr}
    F_\infty^{\mathrm{(n,r)}}(M;\vec\kappa) &= M(- \fs[(n)] + \fsr),
\end{align}\eS%
which, together with the two cases $\fsp$ for $\kappa_m=1$ and $\fsst$ for $\kappa_m=-1$, define the bulk and surface free energy densities discussed in this work.

All considered critical surface free energies can be represented using only four basic free energies,%
\bS\begin{align}
    g^-(\zc) &= -\frac{1}{8\pi}\int_0^\pi\log \big[4 \sin^2\big(\tfrac{\alpha+\varphi}{4}\big)\big] \, \dd\varphi \, , \\
    g^+(\zc) &= -\frac{1}{8\pi}\int_0^\pi\log \big[4 \cos^2\big(\tfrac{\alpha+\varphi}{4}\big)\big] \, \dd\varphi \, , \\
    h^-(\zc) &= -\frac{1}{8\pi}\int_0^\pi\log \big[4 \sin^2\big(\tfrac{\alpha}{2}\big)\big] \, \dd\varphi \nonumber\\
    &= -\frac{1}{4} \log\left[\frac{(1+\zc)^2}{1+\zc^2}\right] \, , \\
    h^+(\zc) &= \frac14 \log\left[-{\zc}_-\right]
    -\frac{1}{8\pi}\int_0^\pi\log \big[4 \cos^2\big(\tfrac{\alpha}{2}\big)\big] \, \dd\varphi \nonumber\\
    &= \frac14 \log\left[-{\zc}_-\right] - \frac{1}{4} \log\left[\frac{1-\zc^2}{1+\zc^2}\right] \nonumber\\
    &=\frac14 \log \left[{\zc}_+\right] \, ,
\end{align}\eS%
with 
limits (we define $L_8 \equiv L^{(0,0,1)}(8,4,-1)$)%
\bS\begin{align}
    g^\pm(\zc^\iso) &= 
    \frac{\log2}{16}-\frac{\catalan}{4\pi} \pm
    \frac{\log\zc^\iso}{16} \pm \frac{L_8}{16} &
    g^\pm(\zc\to1) &= \pm\frac{\catalan}{2\pi} \, ,\\
    h^-(\zc^\iso) &= -\frac14 \log(1+1/\sqrt2) &
    h^-(\zc\to1) &= -\frac{\log 2}{4} \, ,\\
    h^+(\zc^\iso) &= \frac{\log2}{8} &
    h^+(\zc\to1) &= 0\, ,
\end{align}\eS%
from which we can combine all surface free energies,%
\bS\begin{align}
\fs[(o)](\zc)  &= \hphantom{3}g^-(\zc) - \hphantom{3}g^+(\zc) - h^-(\zc) + h^+(\zc) \\
\fs[(n)](\zc)  &= \hphantom{3}g^-(\zc) + \hphantom{3}g^+(\zc) - h^-(\zc) - h^+(\zc) \\
\fsp(\zc)  &= 3g^-(\zc) + \hphantom{3}g^+(\zc) - h^-(\zc) - h^+(\zc) \\
\fsst(\zc) &= \hphantom{3}g^-(\zc) + 3g^+(\zc) - h^-(\zc) - h^+(\zc) \, .
\end{align}\eS%
\CC{
with Hamiltonian limits%
\bS\begin{align}
\fs[(o)](\zc\to1)  &= -\frac{\catalan}{\pi}+\frac{\log 2}{4}
\\
\fs[(n)](\zc\to1)  &= \frac{\log 2}{4}
\\
\fsp(\zc\to1)  &= -\frac{\catalan}{\pi}+\frac{\log 2}{4}
\\
\fsst(\zc\to1) &= +\frac{\catalan}{\pi}+\frac{\log 2}{4}
\end{align}\eS%
where $\catalan = 0.915966\ldots$ denotes Catalan's constant.
}
The surface free energy contributions to \eqref{eq:F∞on} and \eqref{eq:Fon} independent on $\vec \kappa$ are given by%
\begin{align}
\fs^\mathrm{(o,n)}(\zc)&=
\fs[(o)](\zc) + \fs[(n)](\zc) 
    =2\bigl[g^-(\zc)- h^-(\zc)\bigr] 
    .
\end{align}
Finally, the critical bulk free energy density reads%
\begin{align}
    \fb(\zc)&=-\frac{\log 2}{2} 
    -\frac{1}{2\pi}\int_0^\pi\gamma\;\dd \varphi\\
    &=
    -\frac{1}{2\pi}\int_0^\pi\log \left[2\,\frac
    {\sin\big(\tfrac{\alpha+\varphi}{2}\big)}
    {\sin\big(\tfrac{\alpha-\varphi}{2}\big)}
    \right] \, \dd\varphi 
\end{align}
with $\gamma$ from \eqref{eqs:Onsager_c}.
The resulting values for isotropic couplings and in the Hamiltonian limit are given in Tab.~\ref{tab:fs} in the main text.

\end{appendix}





\bibliography{Physik.bib}


\end{document}